\documentclass[12pt]{article}

\usepackage[dvips]{graphicx}
\begin{document}

\title{Dynamics of Internal Models in Game Players\thanks{\sl Submitted to Physica D}
}
\author{{\Large Makoto Taiji}\\
Institute of Statistical Mathematics\\
4-6-7 Minami-Azabu, Minato-ku, Tokyo 106, Japan\\ 
taiji@ism.ac.jp \bigskip \\ 
{\Large Takashi Ikegami}\\
Department of General Systems Studies\\
Graduate School of Arts and Sciences\\
University of Tokyo\\
3-8-1 Komaba, Meguro-ku, Tokyo 153, Japan\\
ikeg@sacral.c.u-tokyo.ac.jp
}
\date{}

\maketitle

\begin{abstract}
A new approach for the study of social games and communications is
proposed. Games are simulated between cognitive players who build the
opponent's internal model and decide their next strategy from
predictions based on the model. In this paper, internal models are
constructed by the recurrent neural network (RNN), and the iterated
prisoner's dilemma game is performed. The RNN allows us to express the
internal model in a geometrical shape.  The complicated transients of
actions are observed before the stable mutually defecting equilibrium is
reached. During the transients, the model shape  also becomes
complicated and often experiences chaotic changes. These new chaotic
dynamics of internal models reflect the dynamical and high-dimensional
rugged landscape of the internal model space. \bigskip

\noindent
Keyword: {\em Iterated prisoner's dilemma game; recurrent neural network; internal model; evolution; learning}\\
PACS: 87.10.+e; 02.50.Le; 07.05.Mh

\end{abstract}

\section{Introduction: Life as Game}

Here we propose a new approach to study the emergence of social norms
and the evolution of communications. The cooperative solutions found
in non-zero sum games are generally unstable. That is, nice players
are replaced by wicked ones before long.  The prisoner's dilemma game
well describes this unhappy situation.  How to bring and to maintain
norms in such a base society has thus been a central issue. From the
game-theoretical analysis to social studies, it has been shown that
norms are maintained by players using a certain set of
strategies. Axelrod has remarked that they should be reciprocal and
nice strategies.  A strategy called Tit for Tat (TfT) carries such
characteristics with the smallest program size\cite{Axelrod}.
However, once TfT strategy dominates a population, it is demolished by
some prepared anti-TfT strategies.  In the other words, no strategy
can be evolutionarily stable in the iterated prisoner's
dilemma\cite{Lorberbaum}.  Since we can never wipe out all base
behavior, imperfect cooperation may be the essential feature of the
dilemma game. On the other hand, we sometimes experience a global
punishment.  For instance, if one breaks a rule, all the members of
his group will be punished as well. From a study of many person's
dilemma games\cite{Axelrod2}, a meta-punishment rule (i.e., a rule does
not only punish one who breaks the law but also punishes those who do
not punish) is required to maintain mutual cooperation. Namely, we
read that some global punishment is required for maintaining social
norms.
 
 The present approach is quite different from the above approaches.
We do not think that social norms are attributed to external
requirements (e.g., to invisible hands).  We rather think they are
attributed to players' internal dynamics. Strategies and punishments
are not forced from without. They should be produced internally by
players in a community.

We often see our daily life metaphorically as game.  But the game we
see as a metaphor is not a set of static game dynamics.  The game
should be dynamical in the sense that game players are taken as
coupled autonomous optimizers with complex internal
dynamics. The necessity of studying the dynamical game was initially
proposed by Rashevsky \cite{rash} and recently emphasized by
R\"{o}ssler\cite{rossler}. The important point in the dynamical game
is not only to find the best (e.g. evolutionarily stable) strategy but
to study how players can deviate from the optimal strategy and how
they perform autonomously.  We argue the complexity
of the game is not attributed to its payoff matrix; we rather 
attribute it to the dynamics of players who can predict the others'
behavior and optimize their own performance.

In the present paper, we prepare a game in advance so that players
don't abstract its rules from situations. Instead, players reconstruct
their opponents' strategy by playing the game. We thus view strategies
as the emergent behavior of players and study the dynamics of internal
images which are constructed within players.

Several researchers have tried to model games based on players with
learning and predicting ability. These models use finite automata to
analyze the effects of memory strategies \cite{aumann}, the procedural
complexity of strategies \cite{rubi}, the relationship between complex
strategies and equilibrium concepts \cite{kalai}, and so on.  In
certain aspects, our approach is very different from those cited
above. We don't represent strategies in finite automata. Instead we
use recurrent networks.  That is, the finite automaton is a limited
class of the network. In our modeling, an infinitely complex finite
automaton can appear in the sense that the automaton can have an
infinite number of nodes \cite{kalai}.  By using this network, we
represent the strategy of an opponent and not of itself; players do
optimize their future moves based on the generated network.

 Representation of the recurrent network structure, which was first
proposed by Pollack \cite{Pollack91}, provides a major advantage over
other models for studying its strategy complexity.  Without counting
the number of nodes of the finite automaton, we immediately notice the
complexity of the strategy by the representation of the network
structure.  Also we discuss the time evolution of those structures
when a game is repeated.  Therefore, a simple IPD game now becomes a
very complicated dynamical system with varying its dimensionality.  We
will show how complicated the dynamics becomes until they ultimately
settle down to an ``evil'' society.  The eventual lost of
norms in our setting implicates that an IPD game is too simple to
sustain mutual cooperation whenever players are cognitive agents.  But
again we have to insist that our main concern here is not to study the
established Nash equilibrium but to discuss the complex transient game
process as a novel dynamical system.

The paper is structured in the following manner:
In the next section we explain the simulation models. The recurrent
neural network, the training method, and the algorithms to determine
next actions will be described. In section 3, the results of
games with fixed strategies are shown. The ability of the recurrent network to learn
deterministic and stochastic automata are confirmed. Then, we
describe the results of games between learning strategies. An origin
of complex behavior and its social implication will be
discussed in section 4.

\section{Model}

In this study we use the IPD game, which involves only two players.
We created two models of game players. The first one is ``pure
reductionist Bob'', who makes the opponent's model by a recurrent
neural network. He thinks that the opponent may behave with simple
algorithms like finite automata, which can be expressed by the
recurrent networks.  The second one is ``clever Alice'', who assumes
that the opponent behaves like ``pure reductionist Bob''. She knows
that the opponent makes her model by recurrent networks and he decides
the next action based on that model of herself. In other words, she
builds the model of herself and treats this model as her image in the
opponent. In the following we describe the behavior of the game
players in detail.

\subsection{Recurrent Neural Networks}

First we explain the recurrent neural network which was used to make
the internal models. Building the model of a player's behavior may
involve many methods such as finite automata, Markov chain, and so on.
In this study we use dynamical recognizers, simple and powerful tools
for studying dynamical behavior from the view points of cognitive
studies.  Dynamical recognizers were first discussed by
Pollack\cite{Pollack91,Kolen94} and have been used independently 
by Elman \cite{Elman} studying the dynamical aspects of language.

Pollack showed that some automata could be learned very well by
dynamical recognizers.  When it could not learn automata, fractal-like
patterns are generated in context spaces. In our simulations these
situation can be considered as incomprehensive or random behavior.

\begin{figure}
\begin{center}
\includegraphics[width=7cm]{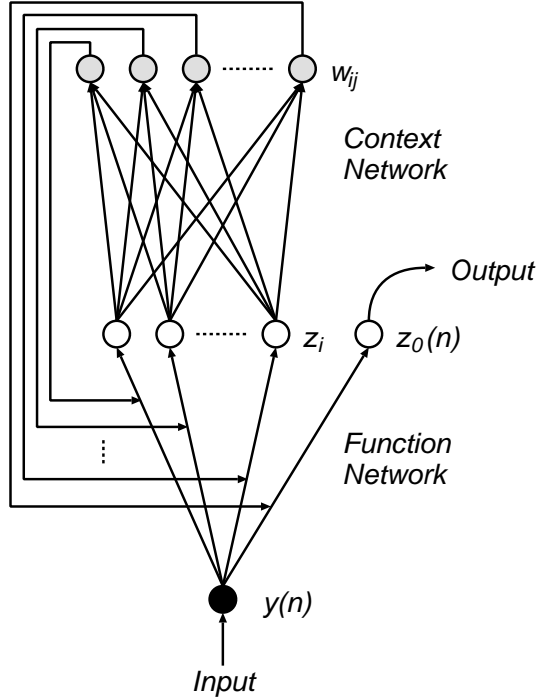}
\caption{Diagram of the recurrent neural network.
Bias connections are ignored.}
\label{fig:rnn}
\end{center}
\end{figure}

Figure \ref{fig:rnn} shows the schematic diagram of the dynamical
recognizer.  The dynamical recognizer is now generally called the
recurrent neural network (RNN).  In particular it is called
``cascaded'' RNN, which consists of a function and a context
network\cite{Pollack91}.  It is quite similar to a two-layer neural
network, though the recurrent outputs are feedbacked not to the input
nodes but to the weights of the function network.  The recurrent
outputs memorize the opponent's current status, and the context
network converts the recurrent outputs to the weights of the function
network which predicts the next action.  Hereafter we call the space
constructed by the outputs from the function network (including both
recurrent and network outputs) as the ``context space''. The output is
taken from a node of the function network.  In this study, only one
input and one output node are necessary since the IPD game has only
two actions, cooperation and defection. We define cooperation as 0 and
defection as 1 in the network. The output is rounded off to 0
(cooperation) and 1 (defection).  The network is expressed by the
following equations.

\begin{eqnarray}
z_i(n) & = & g( w_{i} y(n) + w_{i}^0 ),\nonumber \\
w_i & = & \sum_{j=1}^{N} u_{ij} z_j(n-1) + u_i^b,\\
w_i^0 & = & \sum_{j=1}^{N} u_{ij}^0 z_j(n-1) + u_i^{0b}, \nonumber
\end{eqnarray}

where symbols have the following meanings.

\begin{eqnarray*}
g(x) & : & \mbox{sigmoid function\ } (e^{-x}+1)^{-1}\\
y(n) & : & \mbox{input}\\
z_0(n) & : & \mbox{output}\\
z_i(n) & : & \mbox{recurrent outputs\ }(i=1 \cdots N) \\
w_{i} & : & \mbox{weight of function network}\\
w_{i}^0 & : & \mbox{bias of function network}\\
u_{ij}, u_{ij}^0 & : & \mbox{weight of context network}\\
u_{i}^b, u_{i}^{0b} & : & \mbox{bias of context network}
\end{eqnarray*}

\subsection{Learning}

The recurrent neural network is trained by the back-propagation
method.  In the game, the player knows only his or her own past
actions and those of the opponent. In the case of ``pure reductionist
Bob'', the model of the opponent is built by the recurrent neural
network.  This means that the RNN takes the player's last action as an
input and outputs the opponent's next action. Thus, the target for
training is a series of the opponent's action when the inputs are the
player's actions. However, since the number of training targets
becomes too large as a game proceeds, the weights for learning are
varied for each action so that far actions in the distant past are
forgotten. Thus, the error $E(n)$ after the $n$-th game is given by

\begin{equation}
E(n) = \sum_{k=1}^n \lambda^{n-k} ( z_0(k) - d(k) )^2,
\end{equation}

where $d(k)$ is a target (i.e., the actual opponent's action in the
$k$-th game), $z_0(k)$ is the predicted action by the RNN, and
$\lambda$ is a parameter which controls the memory retention of the
past actions.  Usually we used $ \lambda = 0.9$ for most simulations.
Using the above error we trained the network by using the
Williams-Zipser back-propagation algorithm\cite{Williams89,Doya95d}.
All the weights ($u_{ij}, u_{ij}^0, u_{i}^b, u_{i}^{0b}$) and the
initial recurrent outputs ($z_i(0), i=1 \cdots N$) are adopted based
on the gradients of the error.  The back-propagation processes have
been performed by a fixed count of 200--400, depending on the network
size, to ensure convergence.  We trained the network after each
game. The initial weights are random, and after each game the previous
weights are used as the initial weight for learning.

\subsection{Action}

To determine the player's next action, we use the prediction of the
opponent's future action based on the RNN. First, the algorithm for
pure reductionist Bob is explained. Bob chooses his forward actions up
to $M$ games. Then, he can predict the opponent's actions from his
forward actions, and the expected score can be evaluated. The process
is repeated for all possible strings of actions of length $M$ and Bob
chooses the action with the highest score as the best action. In this
study we used $M=10$ for all simulations.

Clever Alice considers the opponents is a pure reductionist Bob. She
chooses her forward $M$ actions. She predicts the opponent's actions
assuming that he behaves like pure reductionist Bob. Again the process
is repeated for all strings of the length $M$ and she chooses the
action string with the highest score as the best one. In other words,
she predicts her image in the other person and tries to educate him to
have a favorable image through her actions.  In this study we used
$M=4$ for all simulations because of the computational resource.

\section{Results and Discussions}

In this section we describe the games' results and discuss the
observed behavior. At first we briefly describe the results between
the learning strategy and fixed strategies like Tit-for-Tat. The
learning strategy could behave cleverly in the case. Next the results
of the games between the learning strategies (Bob-Bob and Alice-Alice)
are displayed. Through random and complex behavior the systems reach
the trivial fixed point, complete mutual defection. An origin of the
complex transients and the meanings of the final closing with complete
defection are discussed.

\subsection{Games with fixed strategies}

\begin{figure}
\begin{minipage}{8cm}
(a)
\begin{center}
\includegraphics[scale=0.4]{tft-am.eps}\\
\includegraphics[scale=0.5]{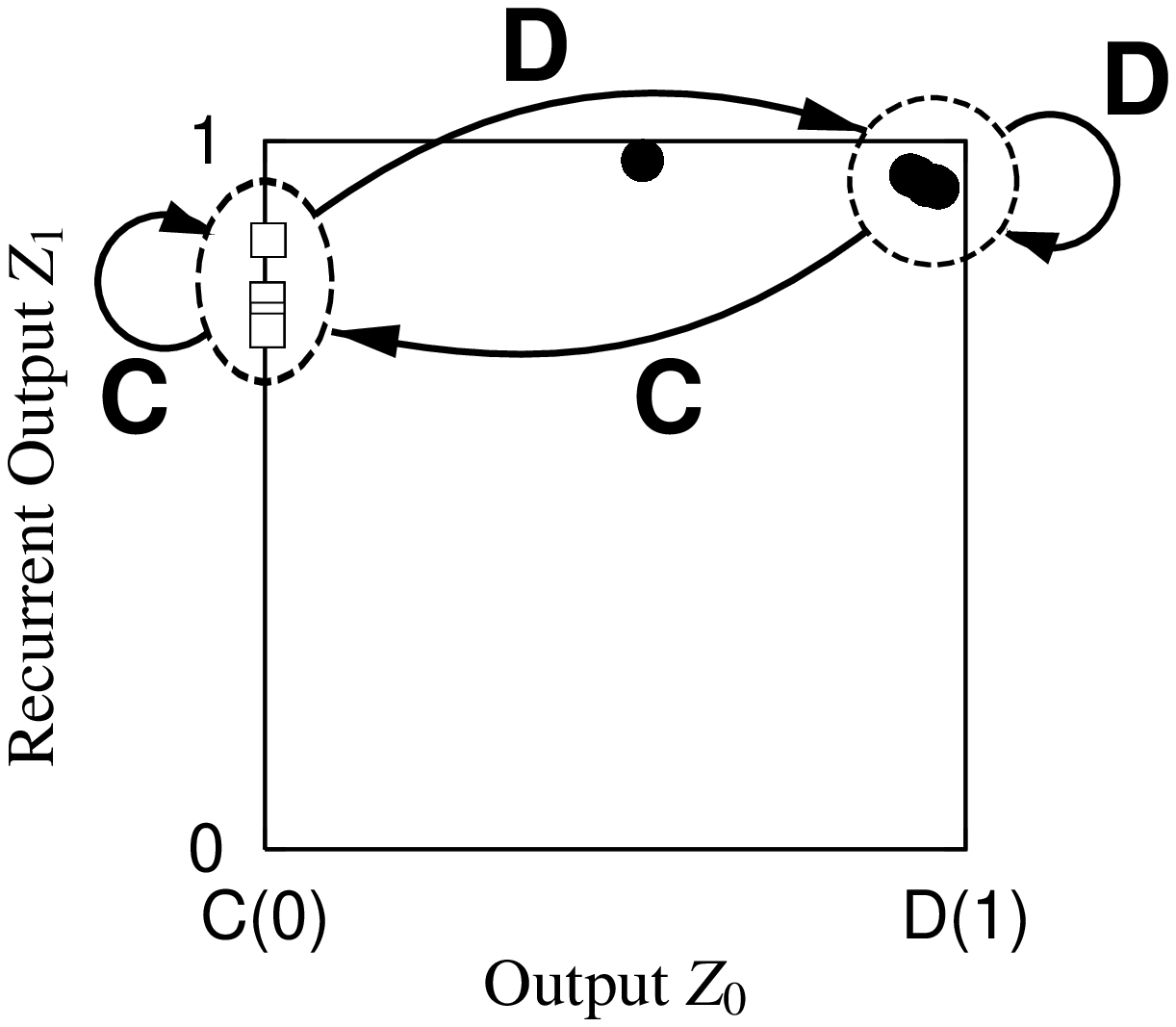}
\end{center}
\end{minipage}
\begin{minipage}{8cm}
(b)
\begin{center}
\includegraphics[scale=0.4]{tf2t-am.eps}\\
\includegraphics[scale=0.5]{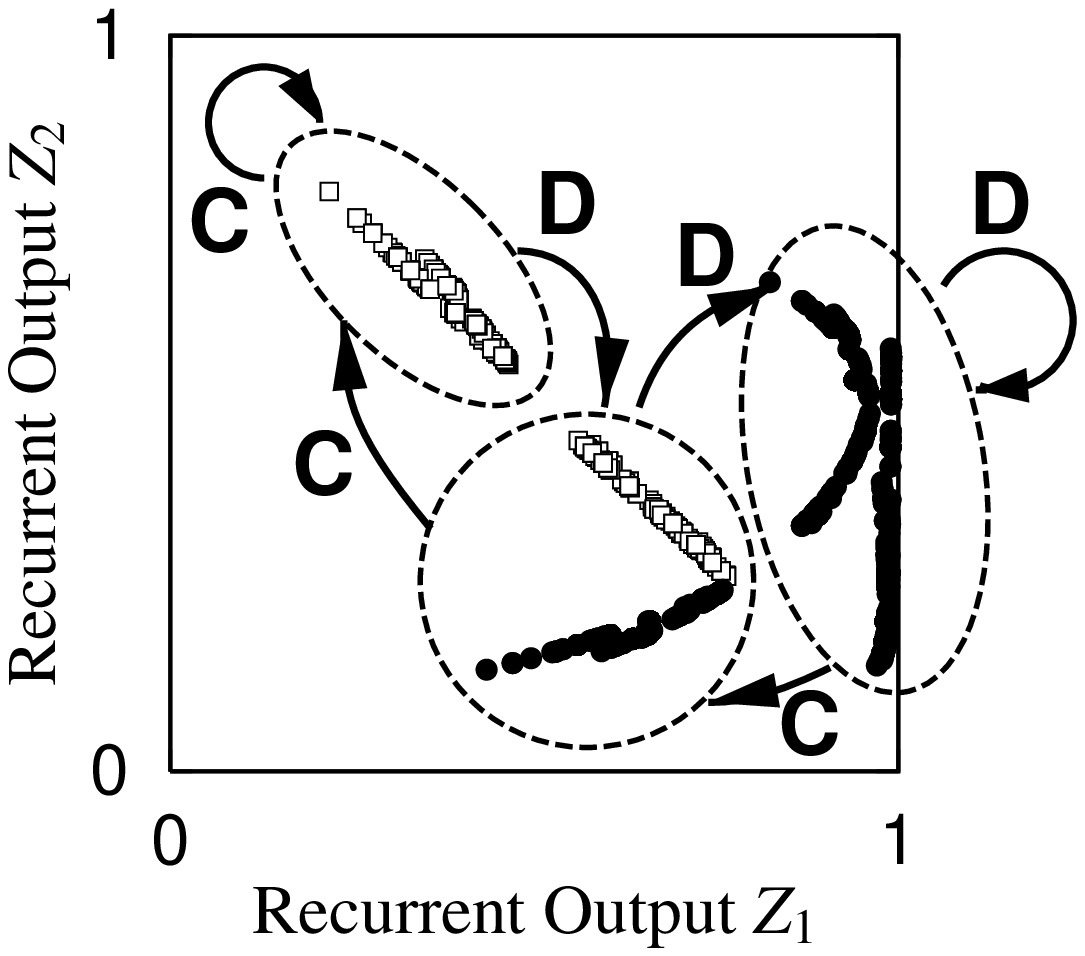}
\end{center}
\end{minipage}
\begin{minipage}{8cm}
(c)
\begin{center}
\includegraphics[scale=0.4]{mixed-am.eps}\\
\includegraphics[width=6cm]{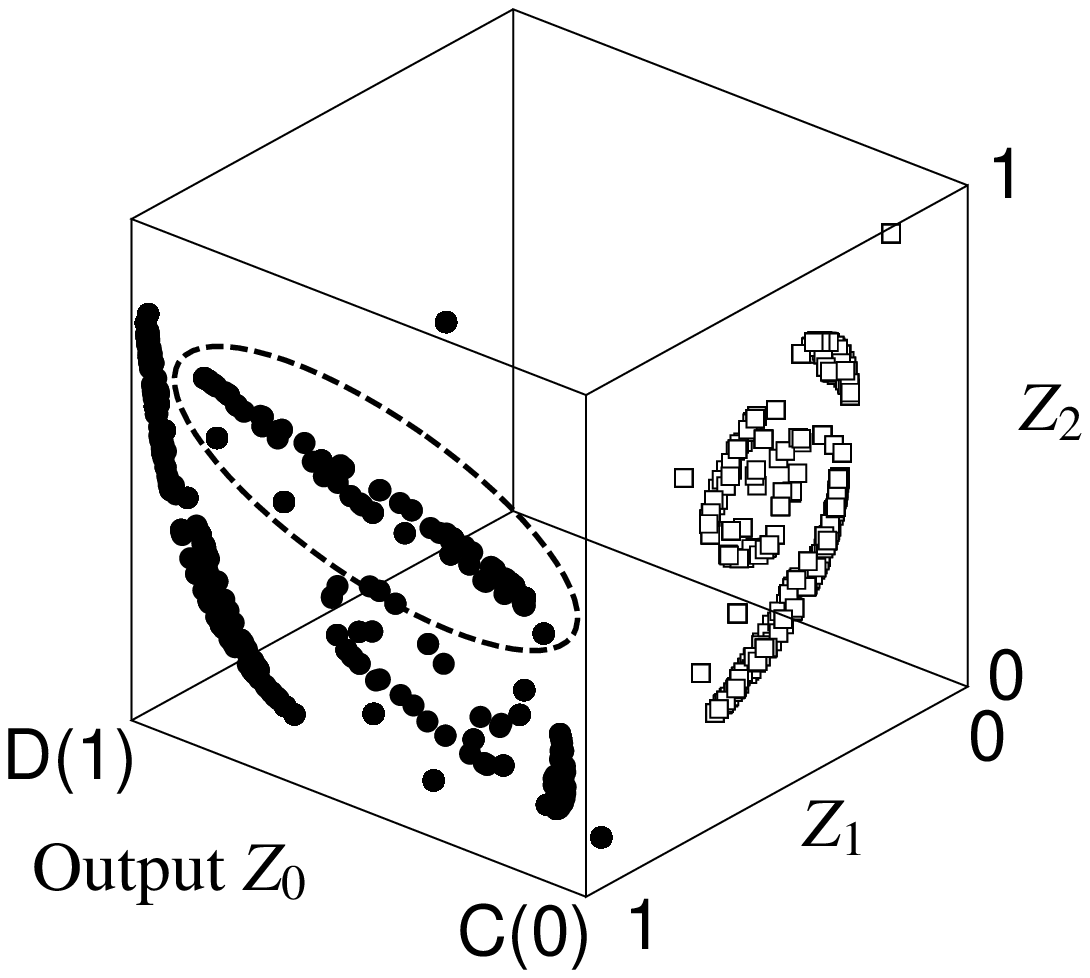}\\
\end{center}
\end{minipage}
\caption{Context space plot of the recurrent neural network after
learning (a) Tit-for-Tat strategy, (b) Tit-for-Two-Tat strategy, and
(c) their mixed strategy. Each strategy is shown in the diagram upon
each graph. Each point represents the output from the function
network with the inputs for all strings of the length 8.  White
squares and black dots correspond to the output after an input of
cooperation and defection, respectively. The networks are shown after
100 games (including the initial 10 games) in (a), 90 games (including
the initial 20 games) in (b) and (c). The learning algorithm could
learn all these strategies and could produce the best actions.}
\label{fig:fixed}
\end{figure}

In the games with fixed strategies, we expressed the strategies of the
opponents by deterministic and stochastic finite automata.
The Tit-for-Tat (TfT), Tit-for-Two-Tat (Tf2T), and their mixed
strategy are used as examples. The first example is the famous
Tit-for-Tat strategy, wherein the player repeats the opponent's last action. 
The structure and dynamics in the context space
after learning is visualized in Fig. \ref{fig:fixed}(a). In the
simulation, the network with two recurrent outputs was used, so the
function network has three output nodes including the target
output. In the figure, two of these three outputs are plotted on the
plane. The points represent the output of the function network $z_i(n)$
for all possible inputs of 8 games. Thus, the figure corresponds to a
transition diagram for a finite automaton.  To learn the TfT strategy,
the random initial action of the player for 10 games {\sl CDDCCCCCDDC}
are given, where {\sl C} and {\sl D} indicate cooperation and defection,
respectively.  The figure is plotted for the network after 100 games
(including the initial games).  The result shows that the network
learned the target automaton (TfT) perfectly. After the 10 games of the
learning period, the player kept complete cooperation for at least more
than 100 games with the forgetfulness parameter $\lambda = 0.95$. The
results do not depend on the initial actions except for very rare ones
such as all cooperation or all defection.

The second example shows the Tit-for-Two-Tat (Tf2T) strategy, which only
defects after the opponent's successive defections. Thus, the Tf2T
strategy is more generous than TfT and requires only one new node
in the finite automata representation.  Figure
\ref{fig:fixed}(b) shows the similar context space plot with a transition diagram.
In this case, also, the network could learn the automaton perfectly.
After the initial action for 20 games {\sl CDDCCCCCDDCDDDCDCCDD}, the
player cooperated for first three games. Then, the player started to
express the optimal action {\sl DCDCDCDCDC$\cdots$}. The network after
90 games is shown in the figure.  We confirmed that the network could
predict perfectly the correct actions of the Tf2T strategy, though the
output $z_0$ is not shown in the diagram.

The third example is the mixed strategy of TfT and Tf2T. When the
player defects, the opponent chooses the TfT and Tf2T strategy with
the probability of 1/3 and 2/3, respectively. 
Figure \ref{fig:fixed}(c) shows the three-dimensional plot of 
outputs from the network.
The initial actions of 20 games are given as

\begin{verbatim}
Player  : CDDCCCCCDDCDDDCDCCDD
Opponent: CCCDCCCCCDDCCDDCDCCC
\end{verbatim}
then the games proceeded as
\begin{verbatim}
Player  : CCCCCCCCCCCCCDCDCDCDCDCCCCCDCCCDCC
Opponent: DCCCCCCCCCCCCCCCCCCCDCDCCCCCCCCCDC
\end{verbatim}
and both players' actions became completely cooperative. Since the original
automaton is stochastic, the network can not learn it
perfectly. However, it could model such random behavior. The
points surrounded by dashed ovals are reached when the transition
occurs from the original cooperative state (shown by a double circle)
by the player's defection.  These points are distributed from 0
(Cooperation) to 1 (Defection) in the output $z_0$. This means that he
cooperates sometime and defects sometime. Thus, the stochastic
behavior can be learned as spread states in the context space. When
the strategy is completely random, the points spread over an entire
region of the outputs.

\subsection{Games with learning strategies}

In this section we show the results when both players have learning
strategies. At first, we describe the games between two pure
reductionist Bobs. To set the initial weights, both players learn the
Tit-for-Tat strategy by playing initially given 20 games.  Actually, both
players' initial actions were set to {\sl CCDCCCDCCDCCCDDCCCCD} and the
opponent was assumed to use the Tit-for-Tat strategy.  
The forgetfulness parameter $\lambda$ was 0.9 so that the players
could forget the initial memories of the Tit-for-Tat.  If the learning
is smooth enough, the new model after leaning will be very close to
the initial model in so far as each player's actions will be consistent with
the model. However, a finite step width in the discrete leaning
process causes the model to deviate from the previous local minima.
Thus, the models could deviate from the initial Tit-for-Tat strategy
and the cooperation could easily be violated.
An example of the players' action is as below.

\begin{center}
\begin{tt}
CCDCCDDCCCCCDDDDCDCCDCCDCCCCDCCDCCDCDDCD\\
CCCCDCCDDDDDDCDDDDDCCDDCDCCCDDDCDCDCDDDC\\
\medskip

CCCDDCDCDDCDCDCCDDCCDCDDCDDDDDDCDDCDDCCD\\
DDDDDDCDDDCDDDCDCDCCDDDCCDDCCDCDCCCCCDCC\\
\medskip

CDDCDDDDCDDDCDDCDDDDCCDDCDDCCDDDDCDDDDDD\\
DCCDCCCCCDDDCDDDDCDDDDDCDDDDCDDCDDCDDDDD\\
\medskip

CDDCCCDCDCCDDDCDDDDDDCCDCCDDDDDDDDDDDCDC\\
DCCDDDDDDDCDCDDDDDDDDCCDCDDCDCCCCDDDDDDD\\
\medskip

DDDDDDCCDDCDDDDDDDDDDDDDDDDDDDDDDDDDDDDD\\
DDCDDDDCDDDDDDDCDDDDDDDDCDDCCDDCDDDDDDDD\\
\medskip

DDDDDDDDDDDDDDDDDDDDDDDDDDDDDDDDDDDDDDDD\\
DCDDCDCDDCDDDDDDDDDDDDDDDDDDDDDDDDDDDDDD\\
\end{tt}
\end{center}

Each row corresponds to each player.  Here the network with $N=4$,
i.e., four recurrent outputs, has been used.  In the earlier games
players tried to maintain cooperation but gradually increased their
frequency of defections. During this transient the series of actions
were very complicated. After 210 games both players reached the
trivial fixed point (Nash equilibrium), all defection.

\begin{figure*}
\begin{center}
\begin{tabular}{cccc} \hline
Game & 0 & 10 & 18 \\
\raisebox{2cm}{Player 1} &
\includegraphics[width=3.2cm]{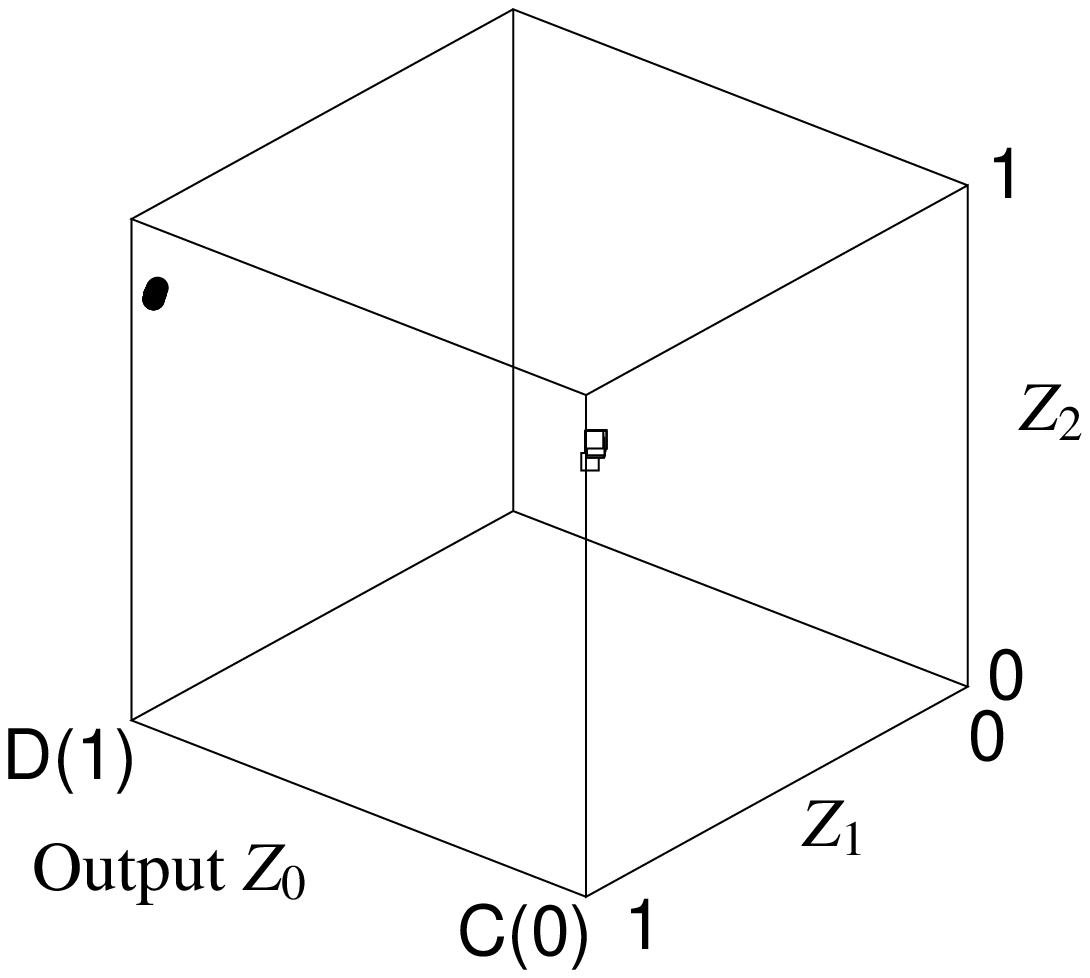} &
\includegraphics[width=3.2cm]{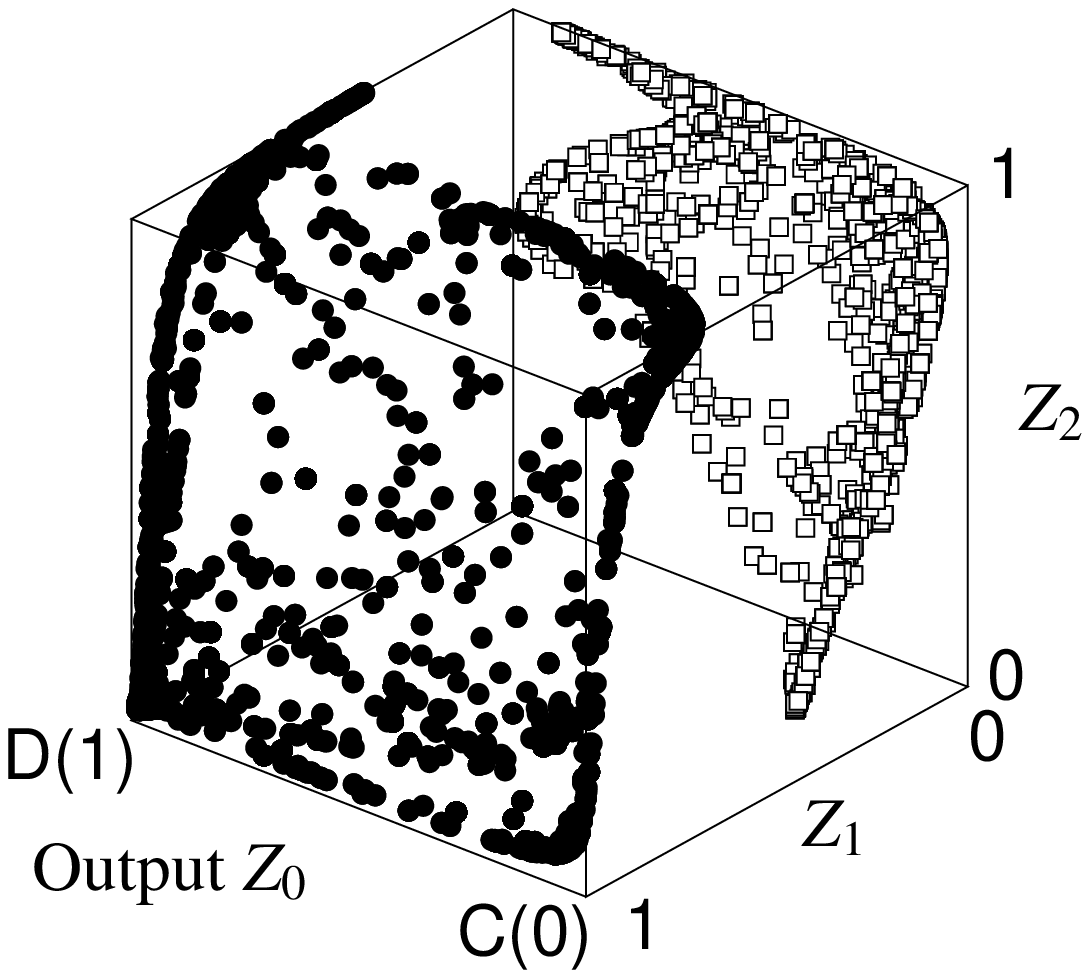} &
\includegraphics[width=3.2cm]{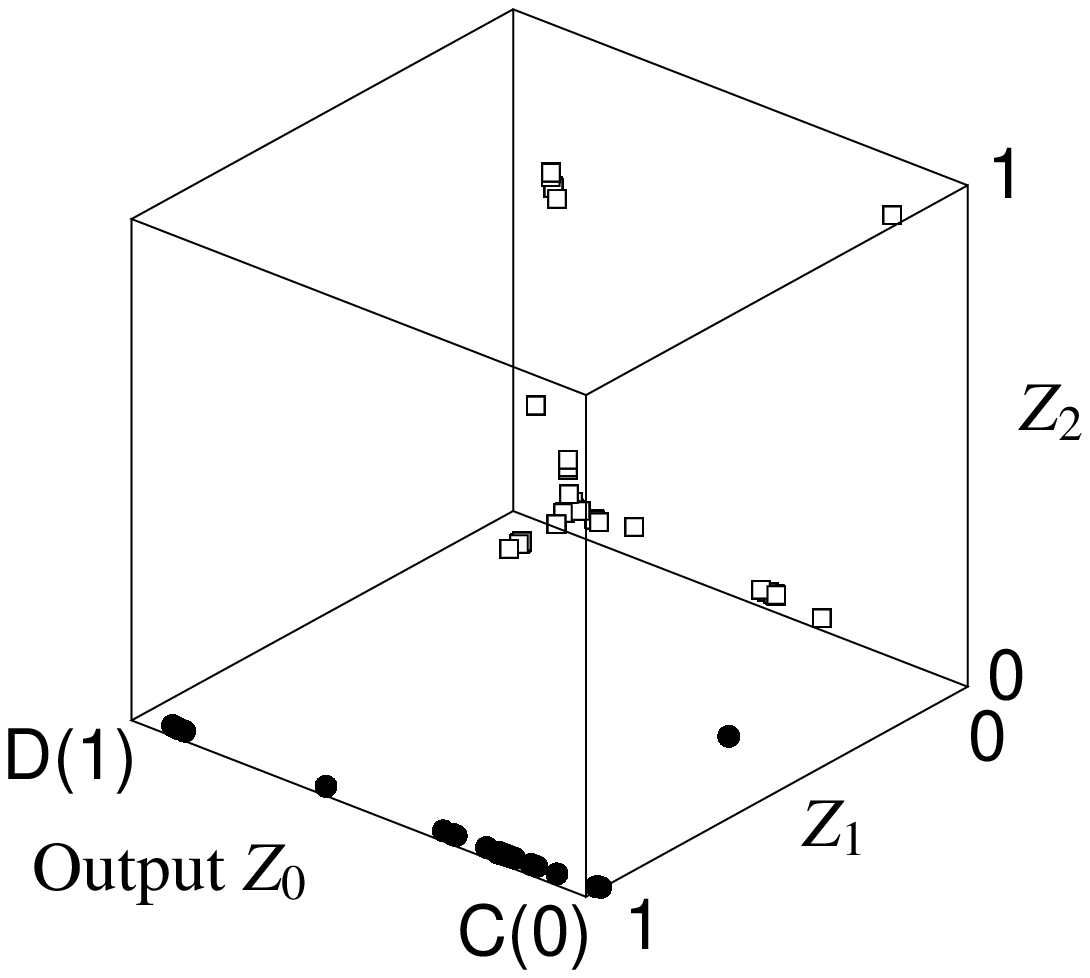}\\

\raisebox{2cm}{Player 2} &
\includegraphics[width=3.2cm]{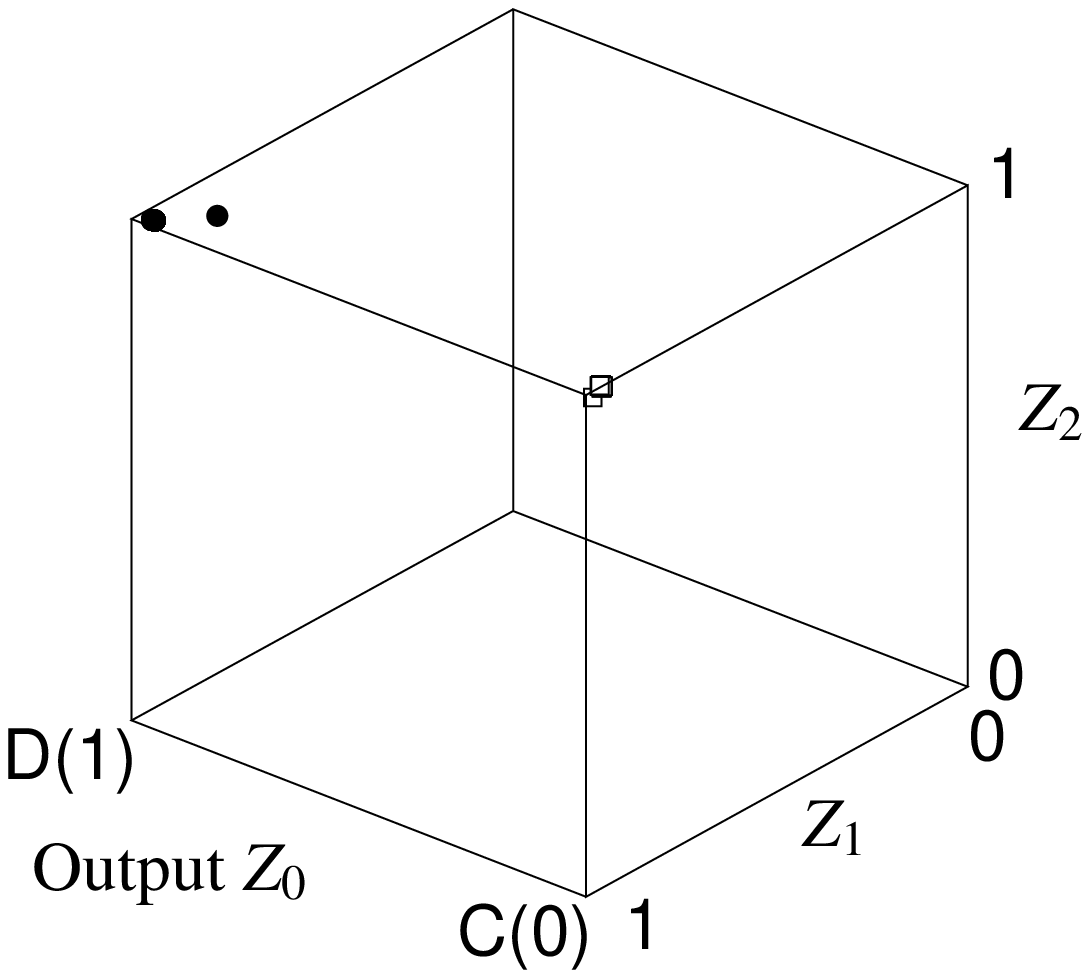} &
\includegraphics[width=3.2cm]{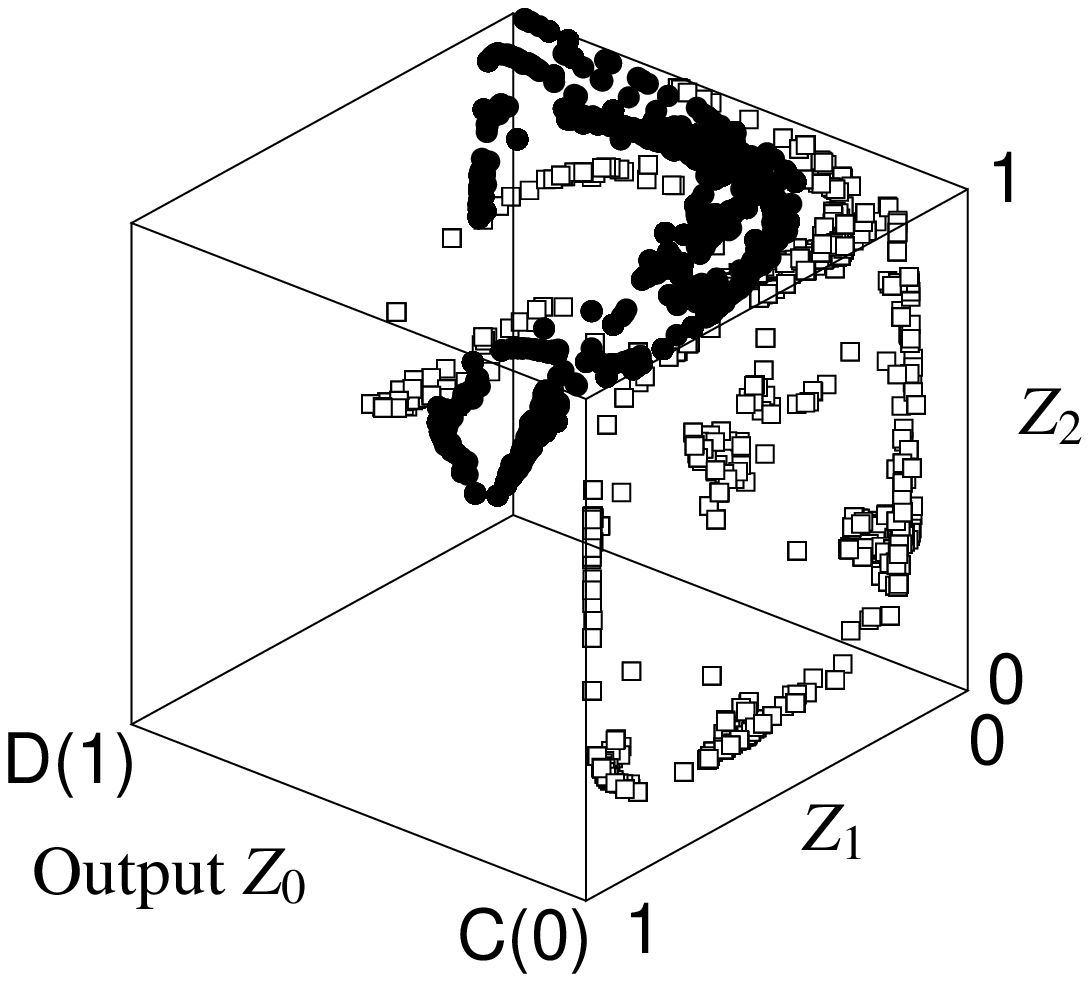} &
\includegraphics[width=3.2cm]{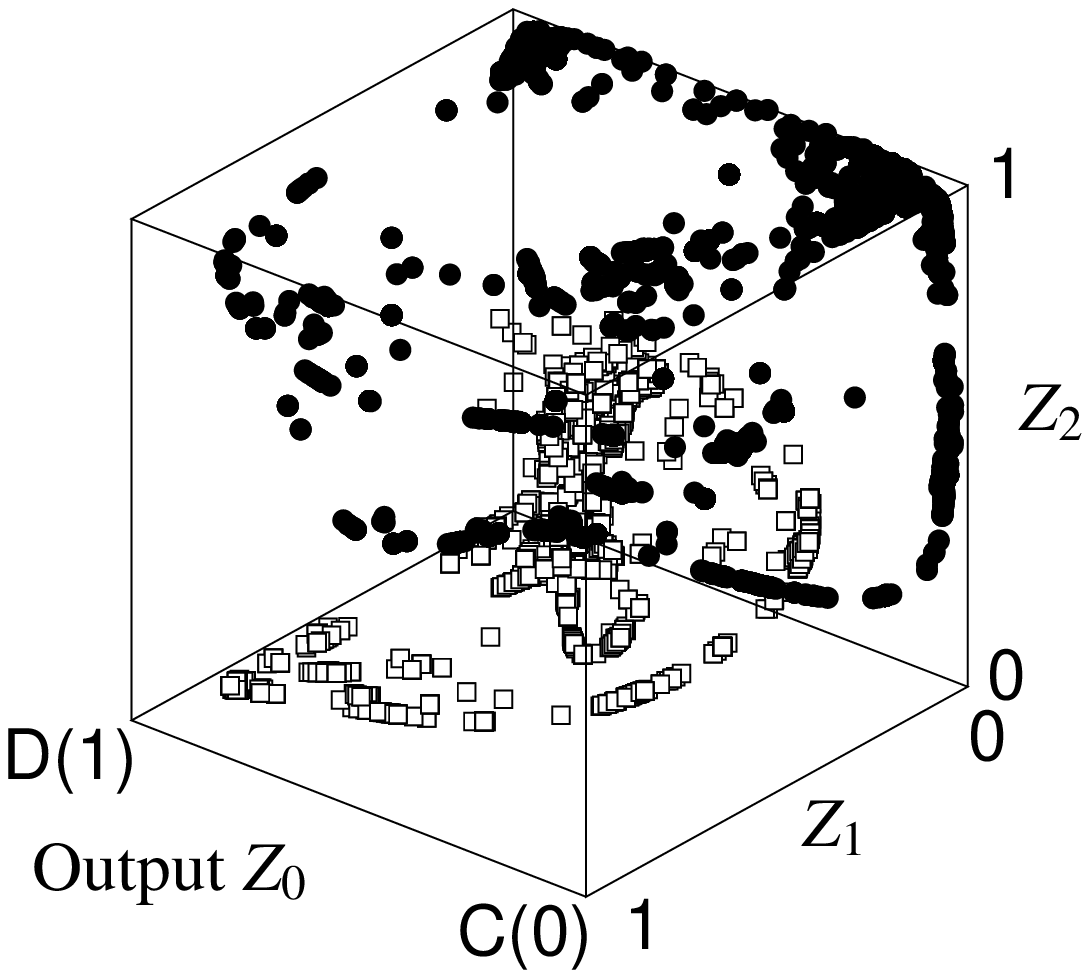}\\ \hline

Game & 20 & 30 & 50 \\
\raisebox{2cm}{Player 1} & 
\includegraphics[width=3.2cm]{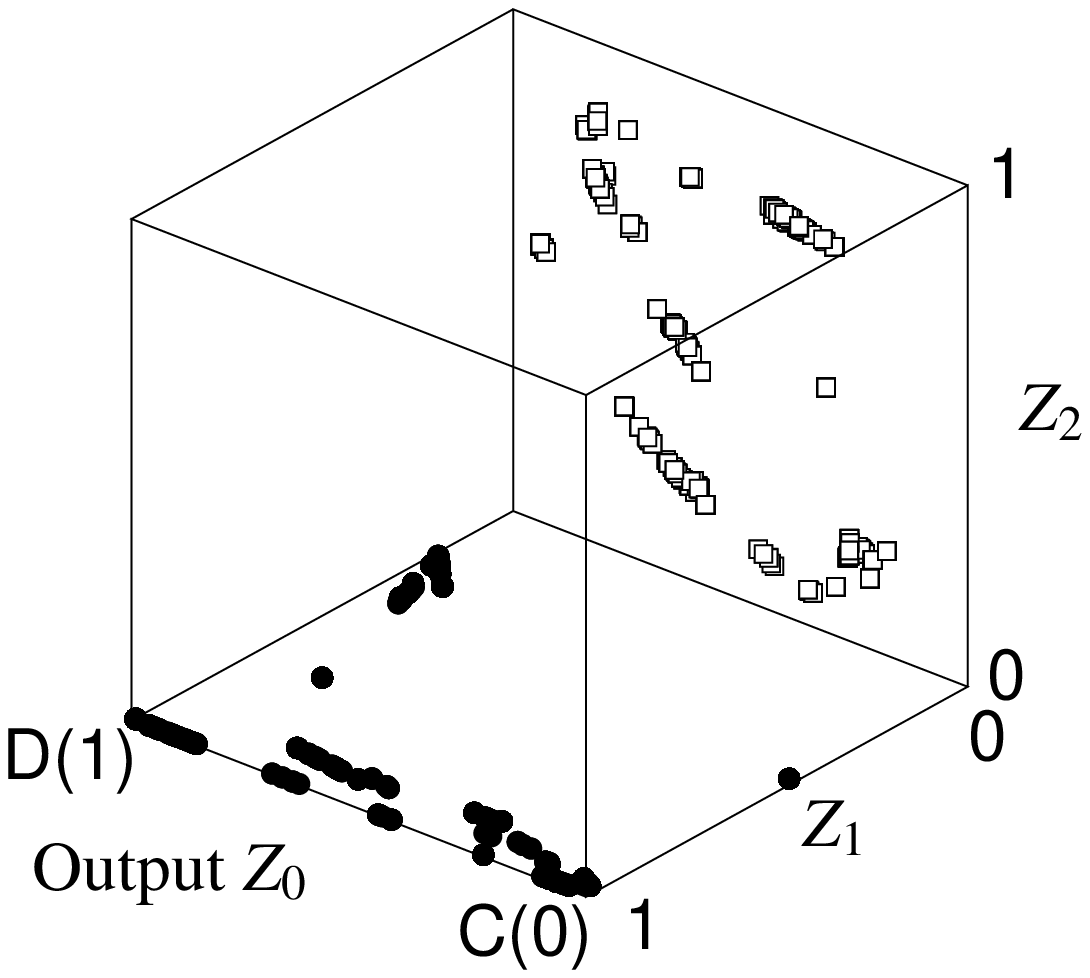} &
\includegraphics[width=3.2cm]{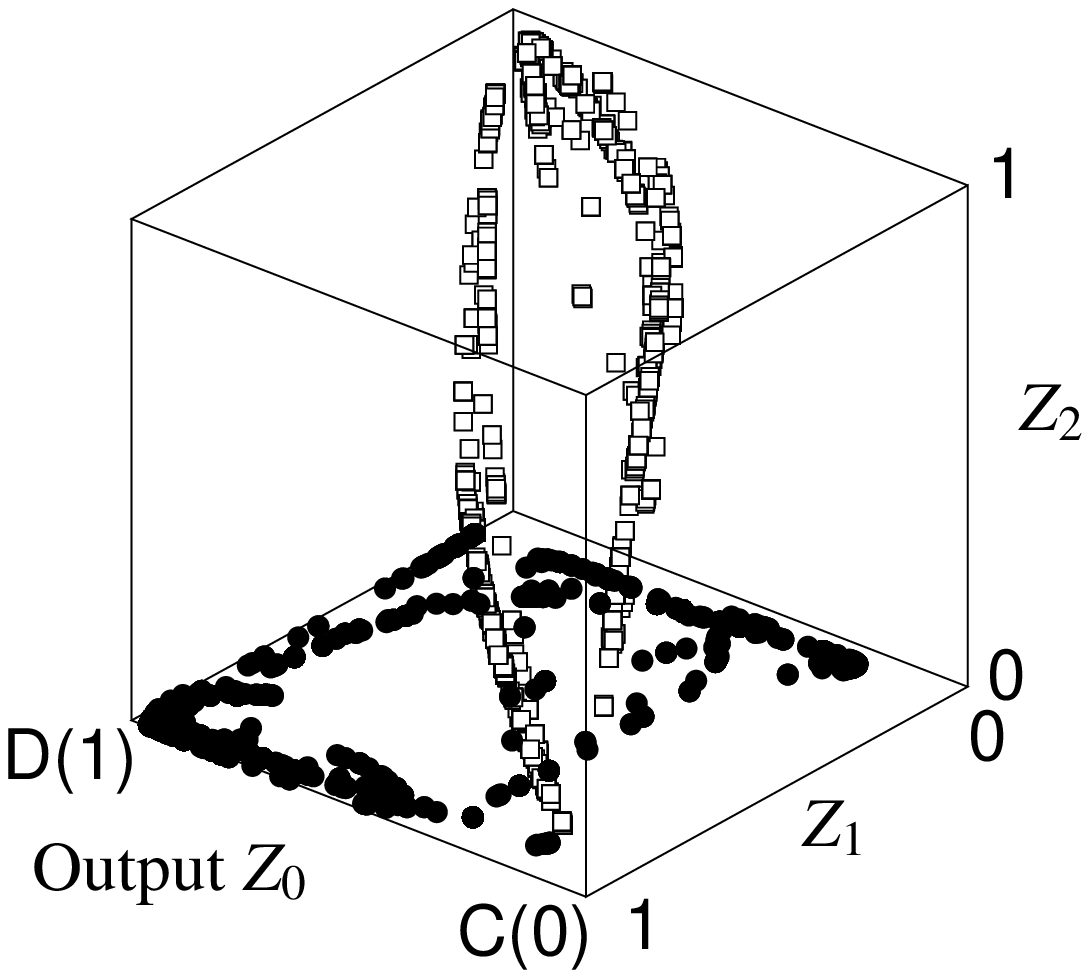} &
\includegraphics[width=3.2cm]{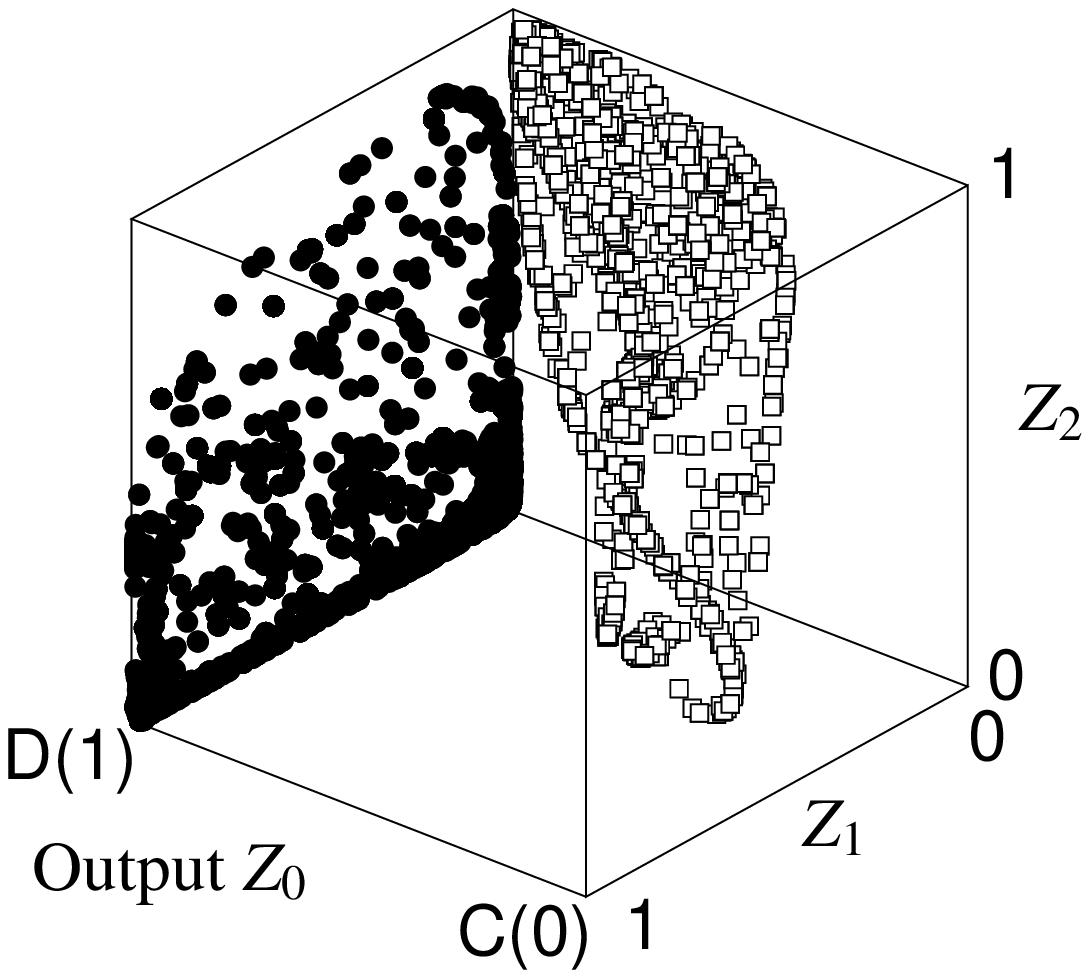} \\

\raisebox{2cm}{Player 2} & 
\includegraphics[width=3.2cm]{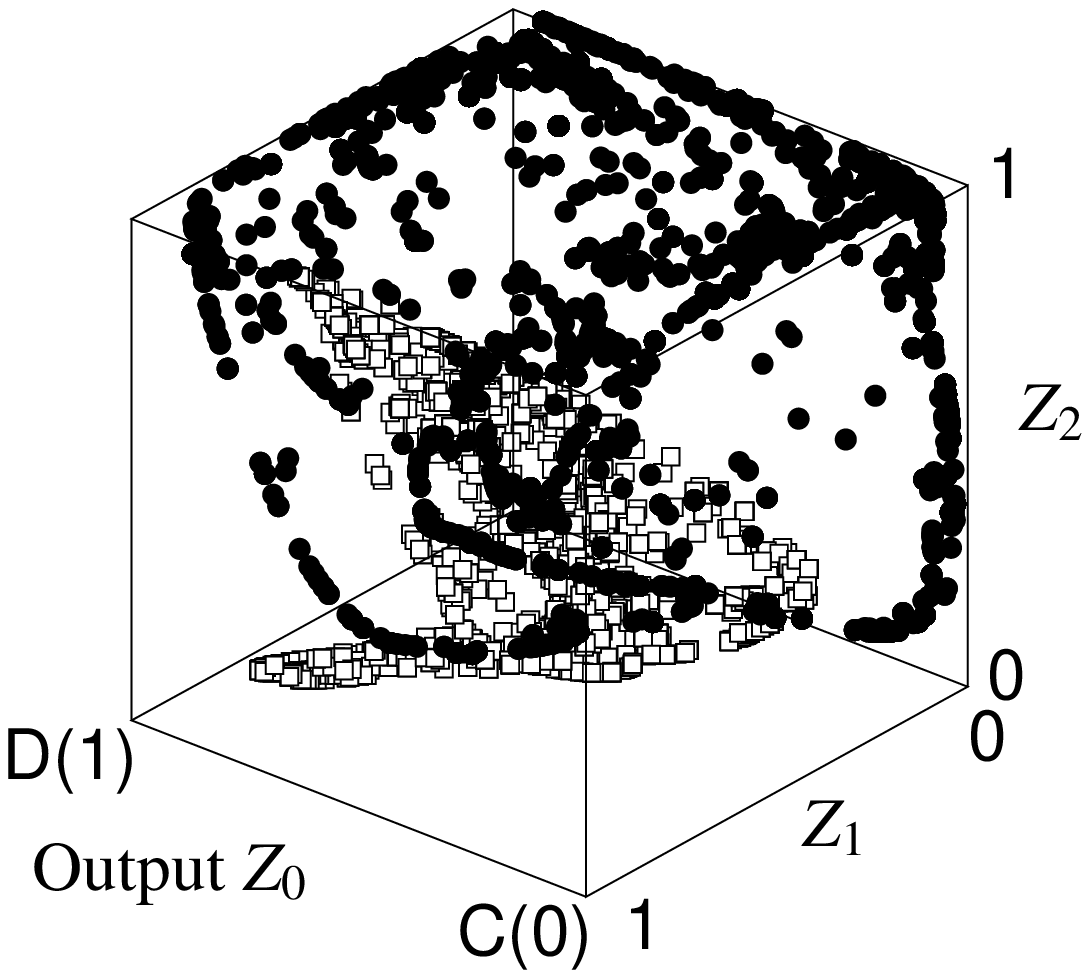} &
\includegraphics[width=3.2cm]{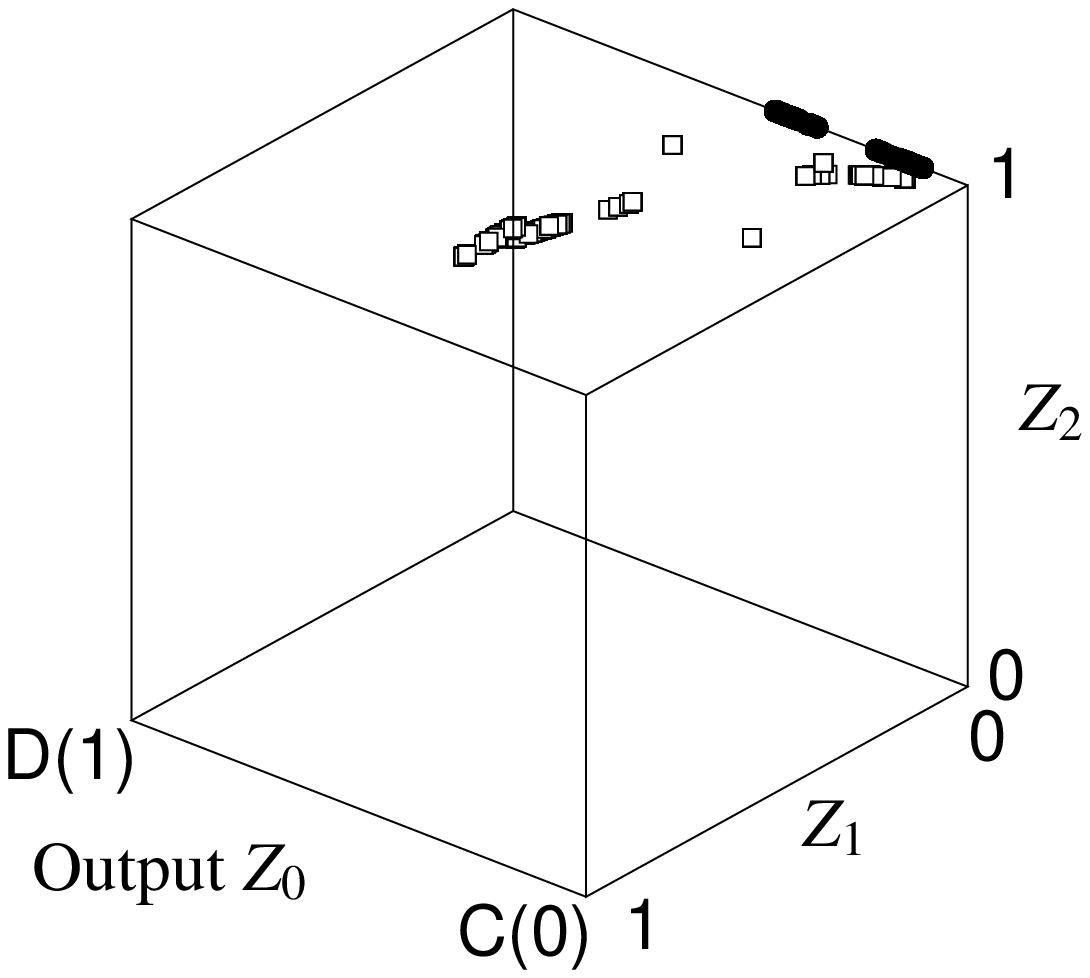} &
\includegraphics[width=3.2cm]{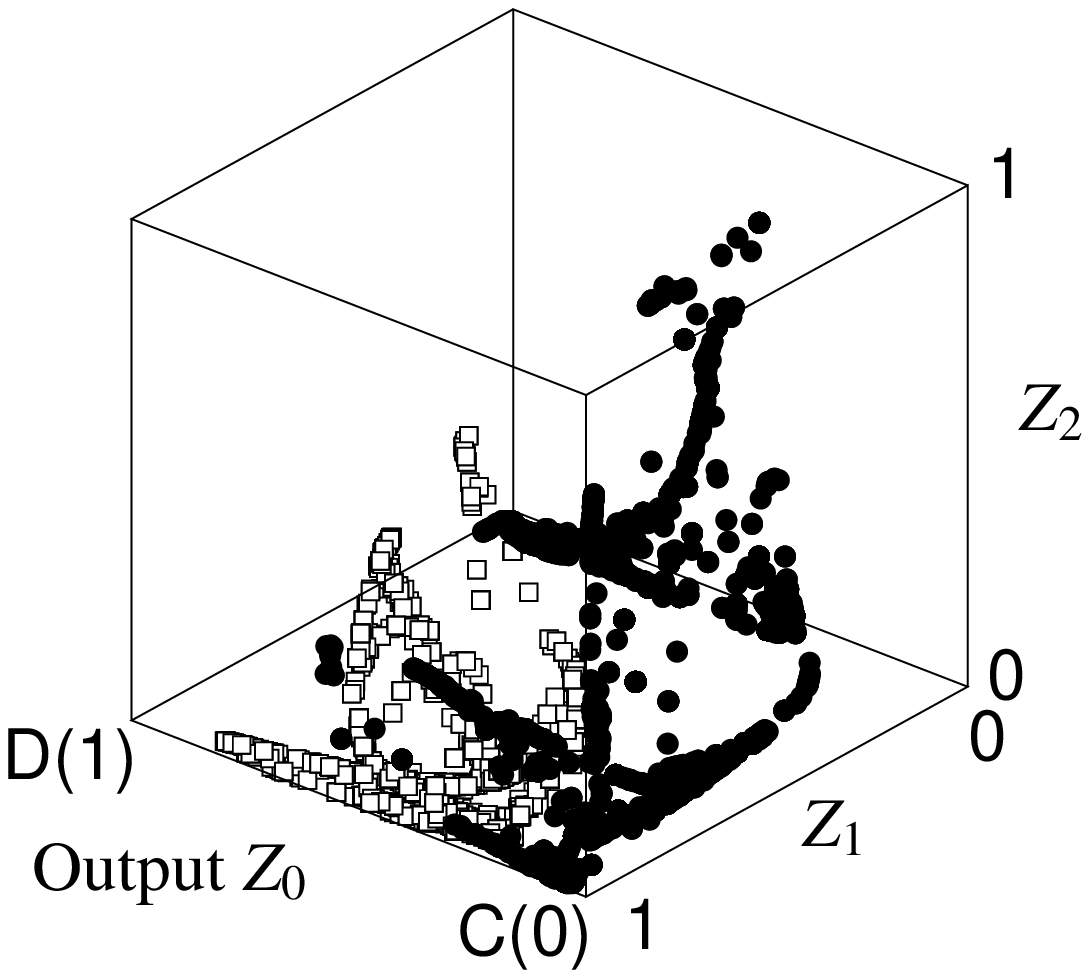}\\ \hline

Game & 100 & 150 & 200 \\
\raisebox{2cm}{Player 1} & 
\includegraphics[width=3.2cm]{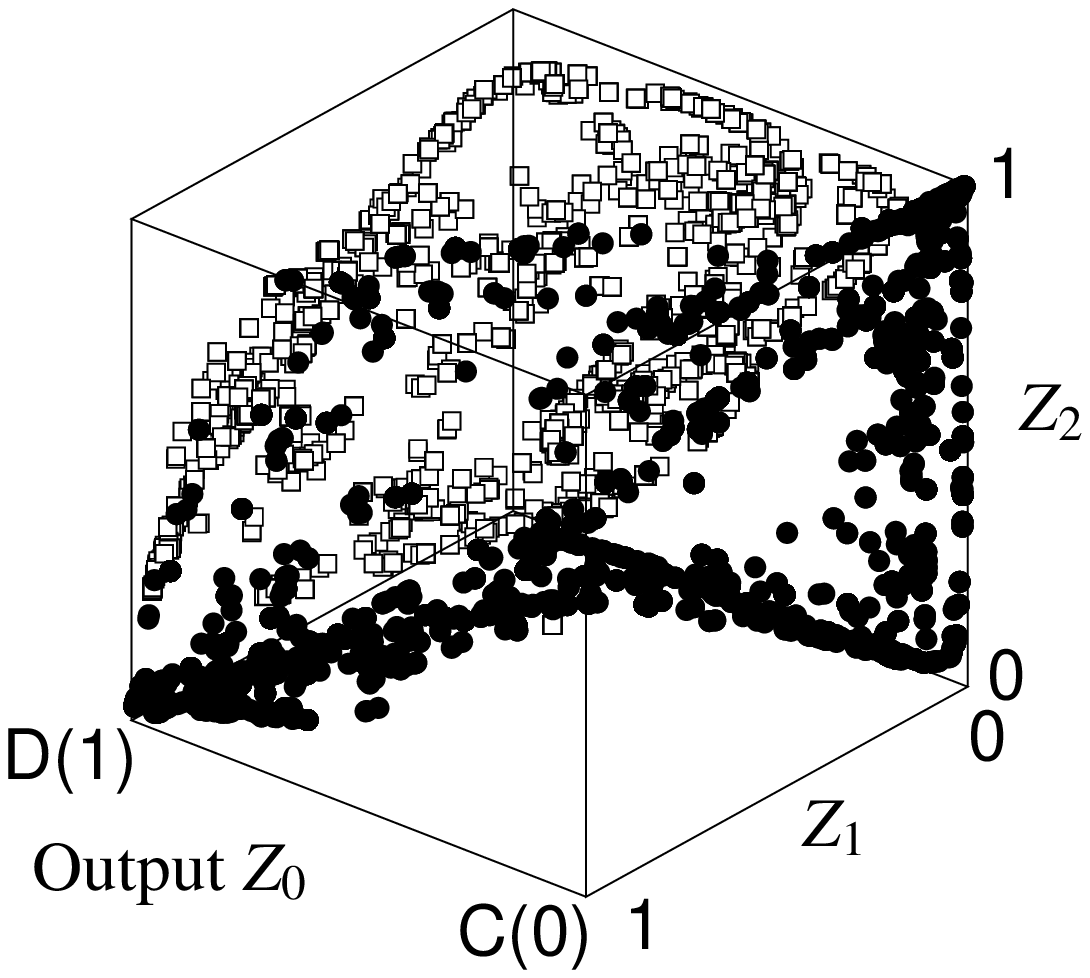} &
\includegraphics[width=3.2cm]{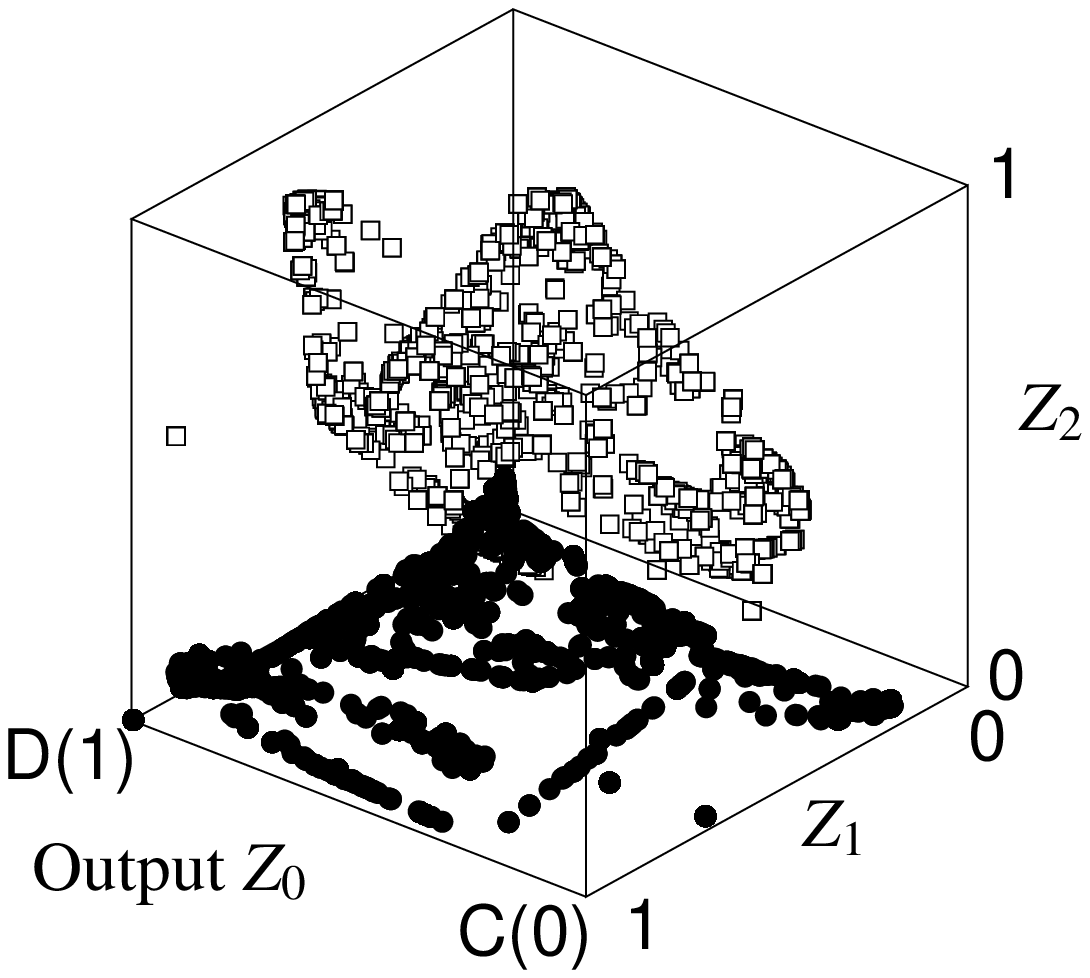} &
\includegraphics[width=3.2cm]{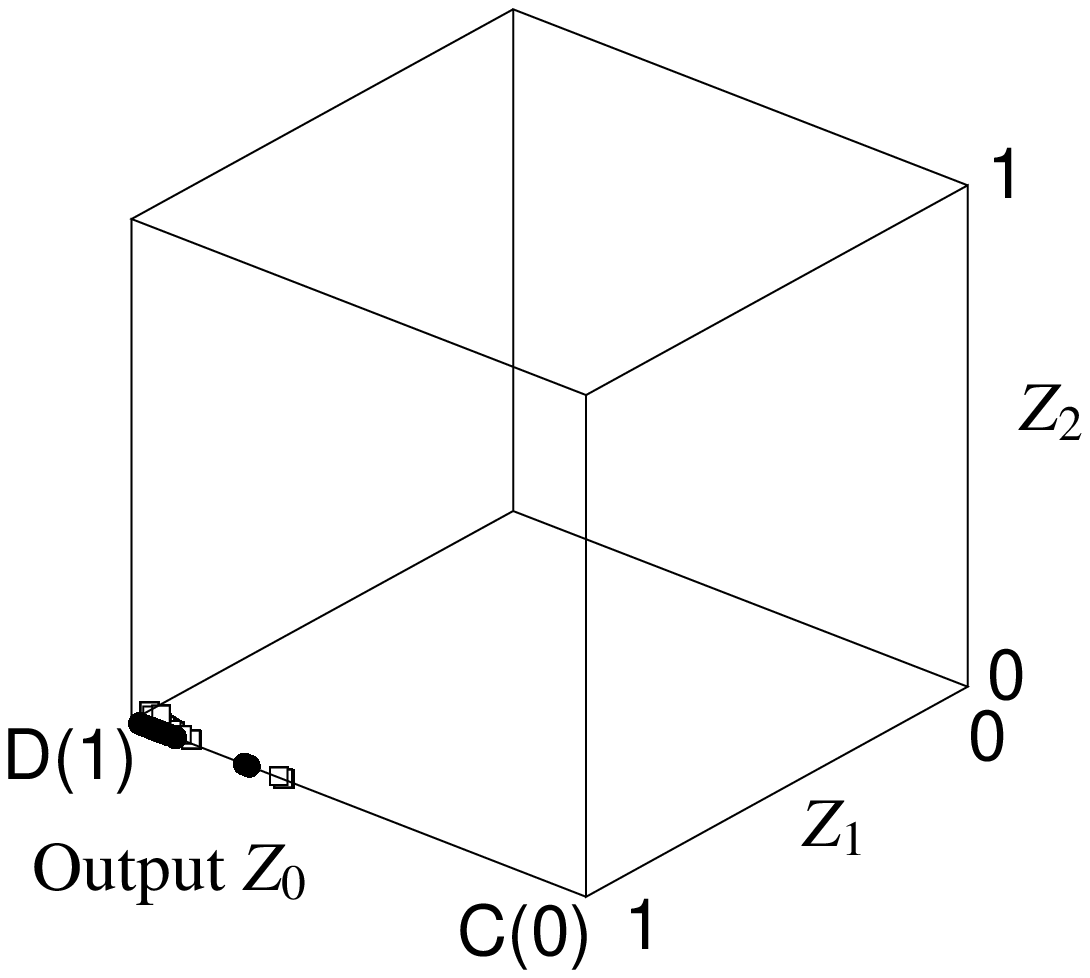} \\ 

\raisebox{2cm}{Player 2} & 
\includegraphics[width=3.2cm]{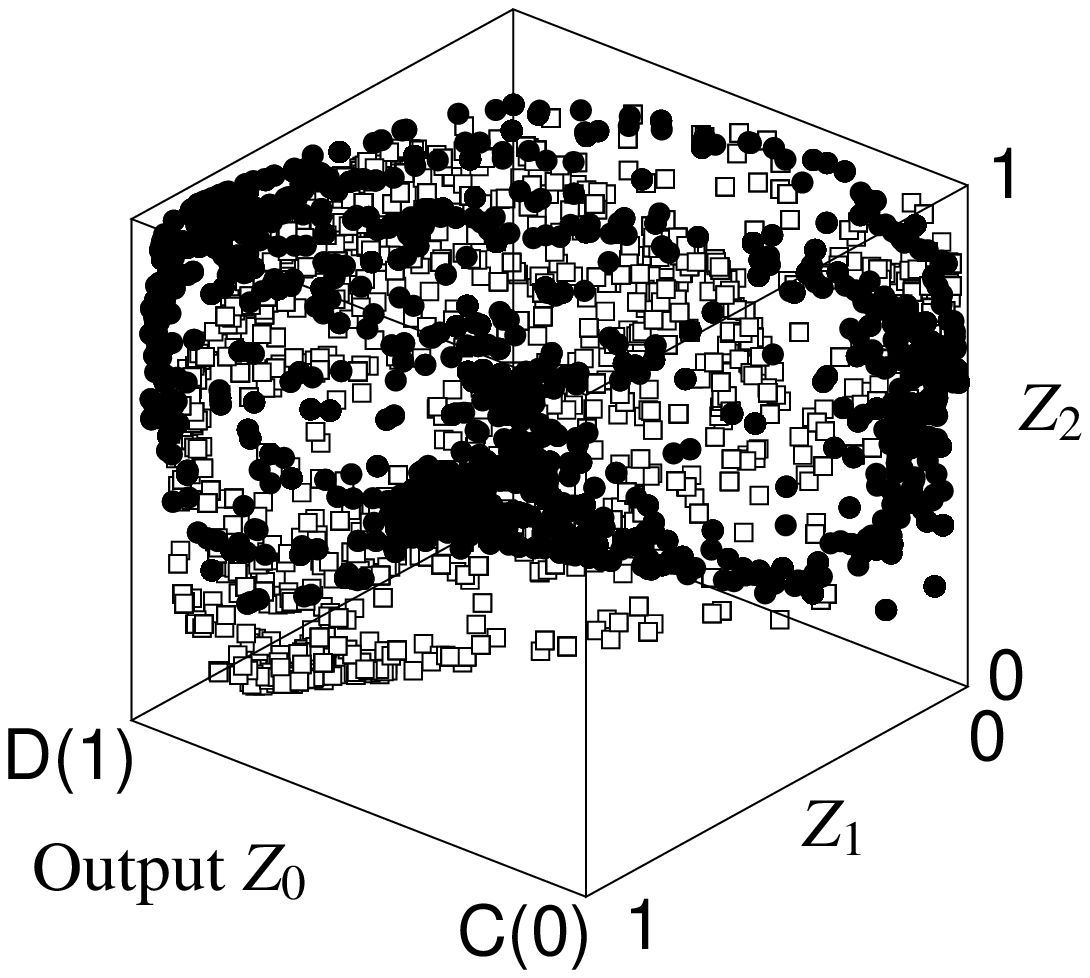} &
\includegraphics[width=3.2cm]{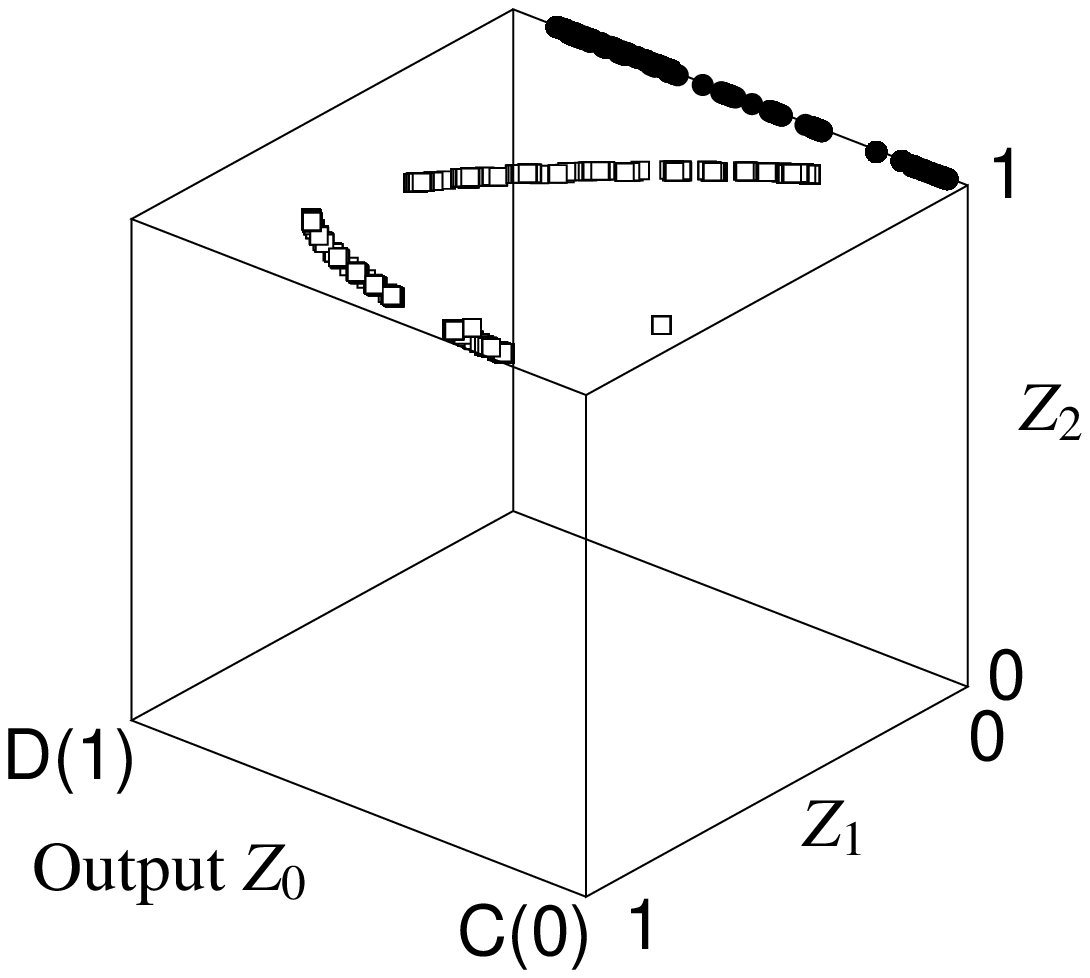} &
\includegraphics[width=3.2cm]{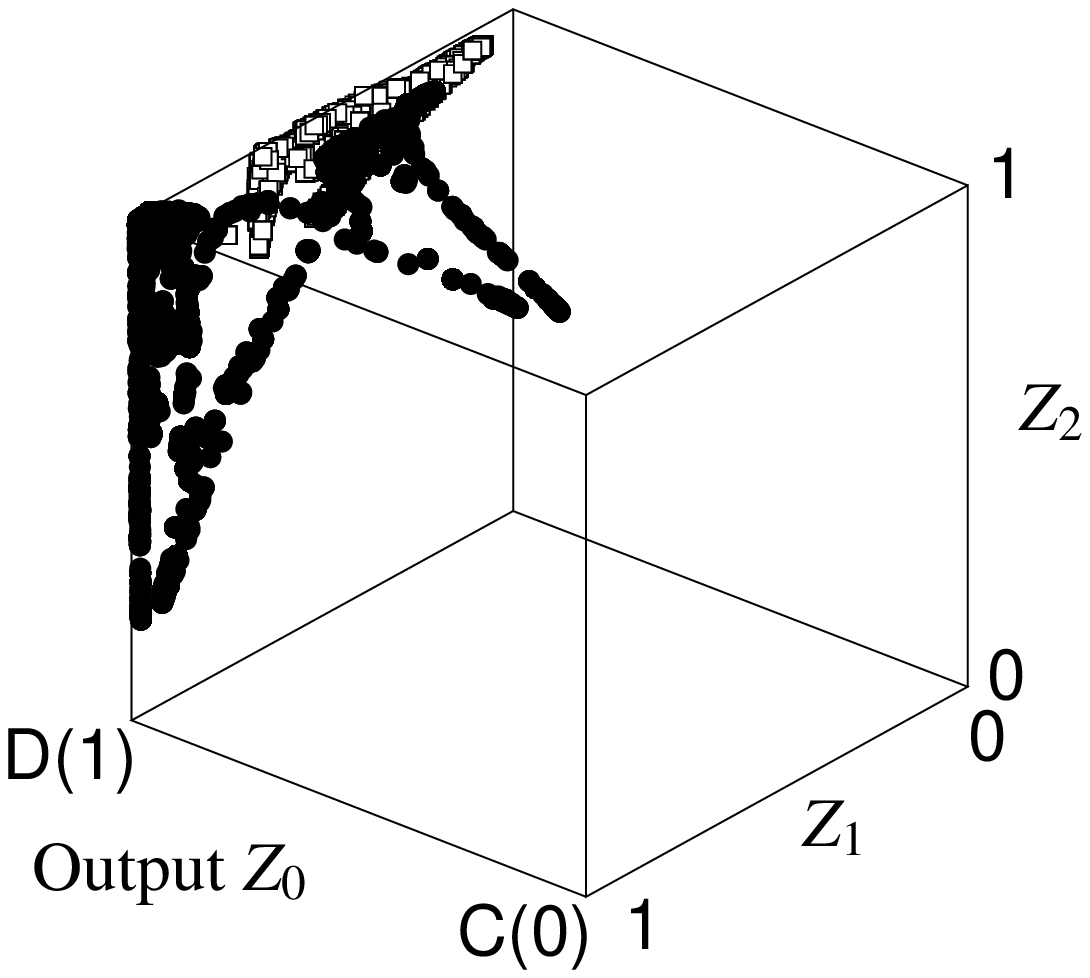} \\ \hline
\end{tabular}

\end{center}
\caption{Context space plot of the recurrent neural network after
games between pure-reductionist Bob players. White squares and black
dots correspond to the output after an input of cooperation and
defection, respectively.  The upper rows correspond to the internal
models of player 2 in player 1, and the lower rows to those of player
1 in player 2. The game generation is shown upon each graph.}
\label{fig:all}
\end{figure*}

Figure \ref{fig:all} shows the context space plot of the internal
models in the players. Only three of the outputs are shown and two of
them are ignored. In the initial game they learned Tit-for-Tat
perfectly, as shown in the upper left corner of the figure. 
Cooperation was broken quite soon, and this internal model became
indistinct after 10 games. Sometimes the models became deterministic,
as seen in the player 1 model after 18 games. However, it
broke again quite soon. In the earlier games each player's model was
rather cooperative, and it gradually became defective as the game
proceeded.  Finally, both players believed their opponents would
always defect.  Thus the system reached the trivial fixed point of 
all defection.

Next we briefly show the result of games between clever Alices who
assume the opponents as pure reductionist Bob. The initial condition
is again Tit-for-Tat; however, in this case this initial condition
means that they believed that the opponents regards themselves as
Tit-for-Tat.  The observed actions are shown below.

\begin{center}
\begin{tt}

DCCDDCDDCDDDDDDDDCCDCDDDDDCCDDCCDDCDDCCD\\
CDDDDDDDDDDDDDDDDDDCDDCCCDDCDDDDCDDDDDDC\\
\medskip

CDDCDDCDDDDCDCDDDDDDCDDDDCDDDDDCDCDCDDC\\
DDCDDDDDDDDDCDDDDCDCDDDDDDDDDDDDDDDDDDD\\
\end{tt}
\end{center}

\begin{figure}
\begin{center}
\begin{tabular}{cc}
\raisebox{3cm}{Player 1} & \includegraphics[width=7cm]{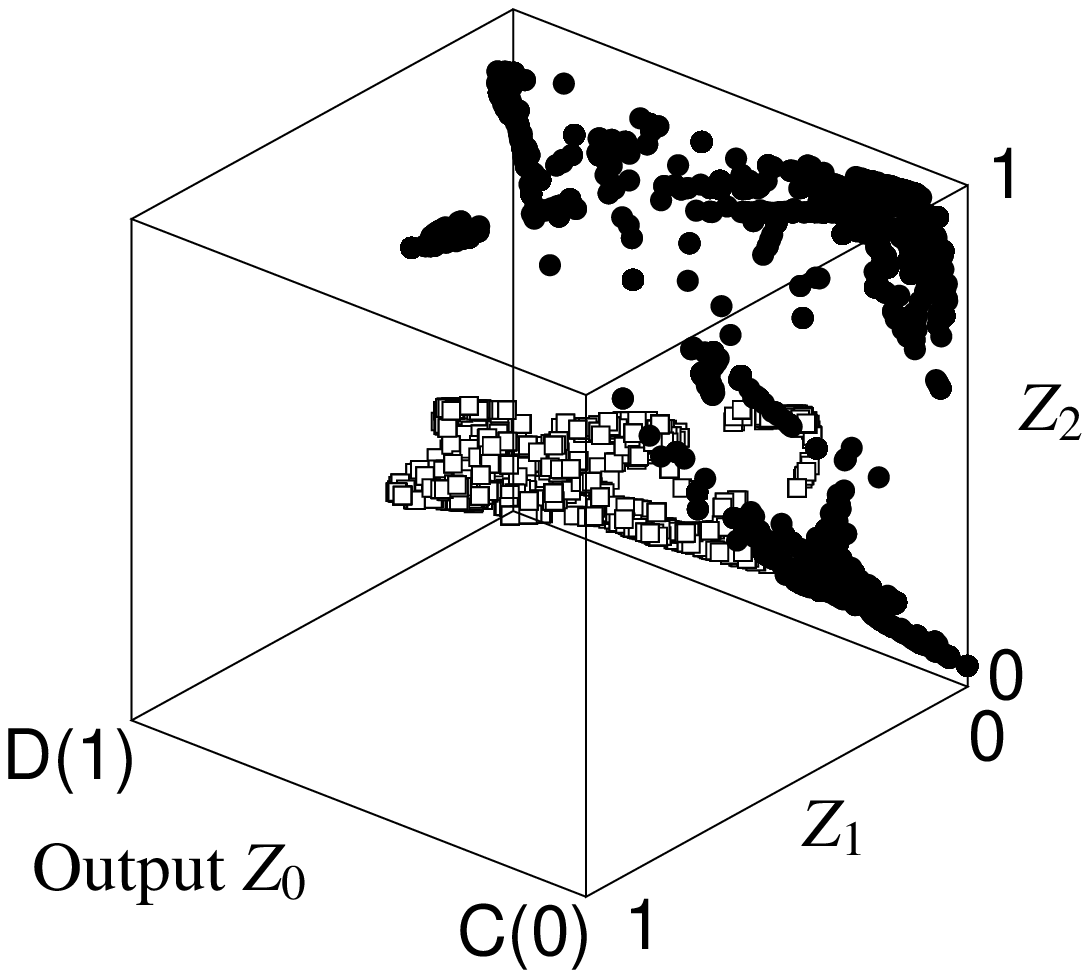}\\
\raisebox{3cm}{Player 2} & \includegraphics[width=7cm]{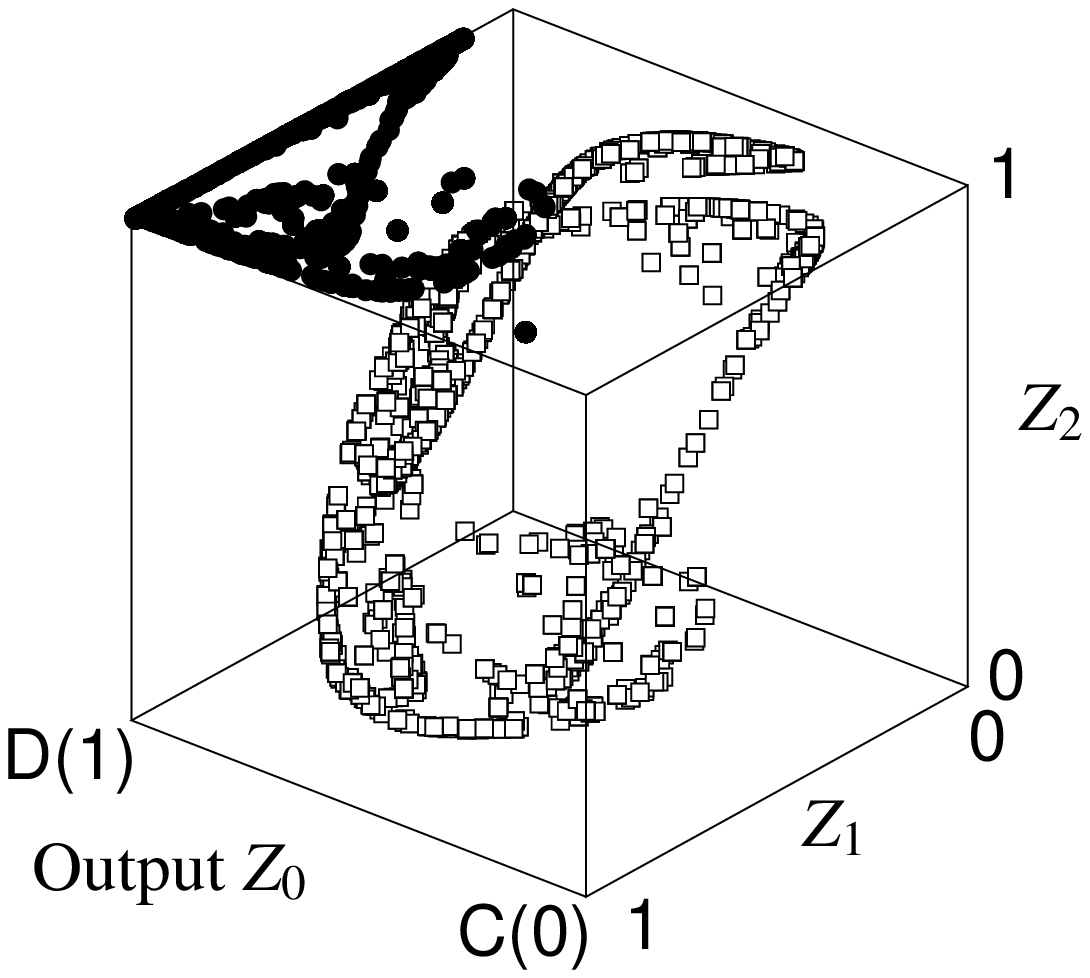}
\end{tabular}
\end{center}
\caption{Context space plot of the recurrent neural network after
41 games between clever Alice players. White squares and
black dots correspond to the output after an input of cooperation and
defection, respectively.  The upper rows correspond to the internal
models of player 1, and the lower rows to those of player 2.}
\label{fig:alice}
\end{figure}

Again the cooperation has been broken quite soon since the player
Alice expects that the opponent (whom she assumes as Bob) will keeps
cooperation as far as he will regard herself as Tit-for-Tat. Thus she
considers that an occasional defection would not break the opponent's
cooperation. However, both players share the identical assumption,
so mutual cooperation is easily violated. Sometimes they cooperated to
educate their opponents to cooperate.  Figure
\ref{fig:alice} shows the context space plot of the internal models in
the players after 41 games. Again, very complicated images were
produced. The internal model gradually became defective as the game
proceeded. Thus each player lost her faith in her opponent's
inclination for cooperation. When she gave up her own attempts to
cooperate, the system reached the trivial fixed point of all
defection.

Why do we see such complex behaviors? In the games with the fixed
strategies, the system could easily reach equilibria. However, in the
games between learning strategies, complicated transients were
observed. The length of the transients, i.e., the length of time it
took to reach the fixed point of complete defection, depends on the
size of the networks. When the number of the recurrent output nodes
was 2, equilibrium was reached after 90 games. When the network
size is small, the player tends to over-simplify the opponent and
easily recognizes his strategy as all-defection.  As the network size
becomes larger, the ability to express complex models increases. Thus
the transients become longer.

\begin{figure}
\begin{center}
\includegraphics[width=10cm]{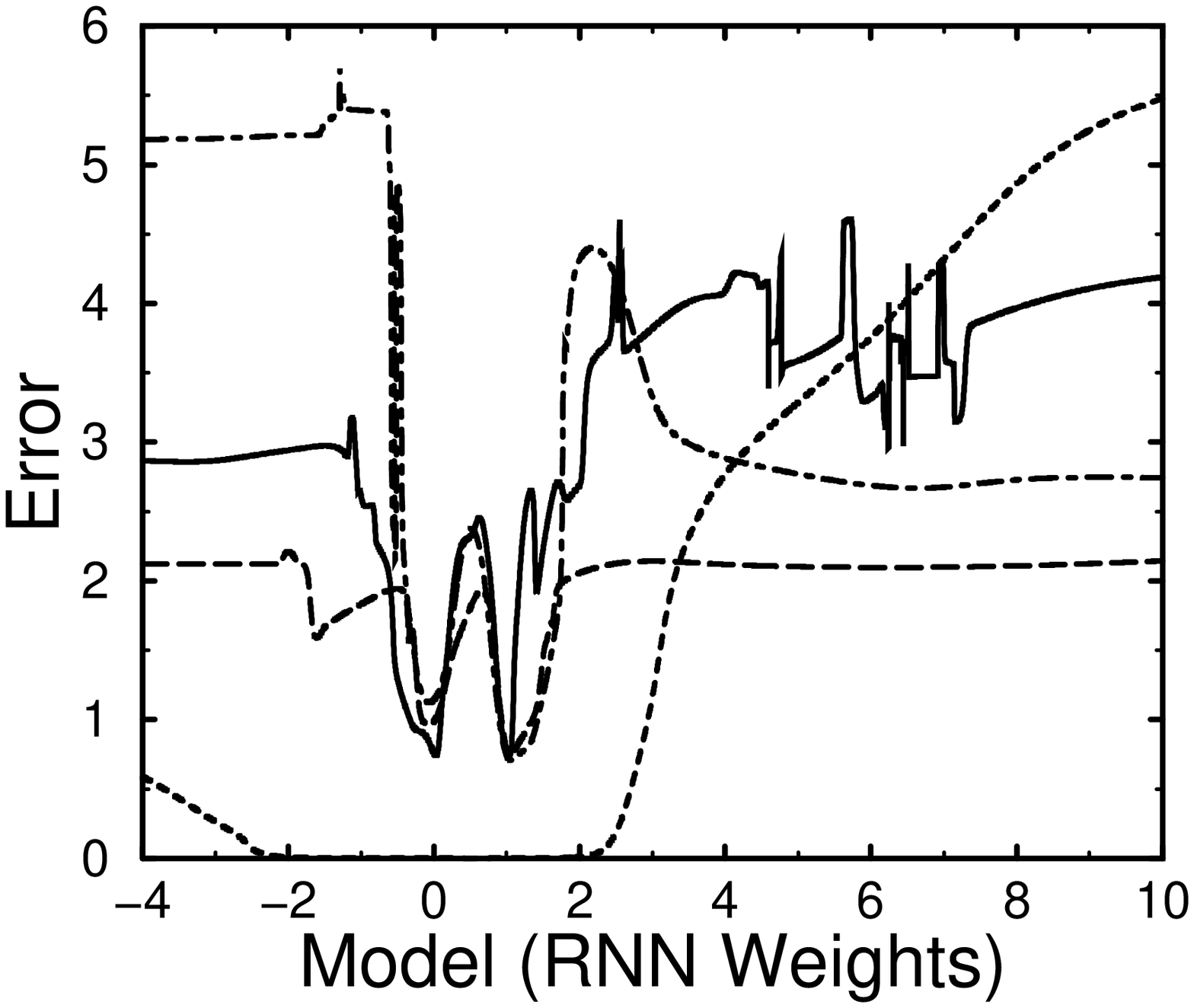}
\end{center}
\caption{Rugged landscape of the error $E(n)$ in the model space.
A one-dimensional line in the space on two local minima is taken as the
horizontal axis. The minima are normalized to 0 and 1. All the network 
weights are linearly interpolated or extrapolated on the axis.
Dashed line: after game 0; dot-dashed line: after game 10;
solid line: after game 20; long-dashed line: after game 180.
}
\label{fig:landscape}
\end{figure}

Figure \ref{fig:landscape} shows an example of the error landscape in
model space.  Since there are $4N^2+2N$ parameters in the model,
it is difficult to show them all. In the figure only models on a
line in the 72-dimensional space are presented. The line was
determined to pass two local minima, which were obtained by creating
internal models from two different initial weights. These minima
correspond to 0 and 1 in the horizontal axis, and all parameters are
linearly interpolated and extrapolated on the line between these
minima. The error $E(n)$ are calculated and plotted for each
model.

In the initial stage when each player successfully believes that the
opponent is Tit-for-Tat, the landscape is smooth and has only one
local minima. However, after a few ten steps it becomes quite complex
and reveals a lot of local minima. At the games' end when each player
assumes that the opponent always defects, again it gradually returns
to a simple flat landscape. These complex landscapes (Fig. \ref{fig:landscape})
  reflect this complex behavior. 
The error increases after each new game,
then the model changes to a new local minimum when the error exceeds
the barrier. Thus such an itinerancy between local minima has been
observed, and constitutes an origin of complex behavior.  Note that these
landscapes do not indicate that the evolution of the behavior is
considered the optimization process in the landscape.  Each player
assumes that the landscape reflects the reality of the opponent, but
actually it is almost always an illusion.  Since no unique model has
an ability to express behavior, the landscapes have many local
minima with similar errors.

A strategy based on internal models can not be expressed by the
strategy itself. Pure reductionist Bob actually has no fixed strategy,
so he can not make a model of himself. He changes his strategy every
game based on his opponent's actions. Without being aware of it, they
change their opponent's strategy.  Clever Alice can make a model of
Bob, though she can not make a model of herself. To make a model of
themselves, an infinite regression will occur. Thus, with learning
strategies players can never completely understand others and behave
rather randomly with no definite strategy except during the stable
equilibrium of the state of all defection.

The misunderstanding of others and semantic complexity in the model
space are essential to observe complex behavior like in games between
learning strategies.  We think that this scenario is generic and not
specific to the RNN. Morimoto and us performed similar simulations
using finite-state automata to build internal
models\cite{morimoto}. Again, very long and complex transients were 
observed.

\subsection{Games between learning and double-learning strategies}

Next, we discuss the games between the learning and double-learning
strategies. In this case, the players can maintain collaboration for some
time; ultimately, however, the equilibrium of all-defection arose.
The observed actions are shown below.

\begin{figure}
\begin{center}
\includegraphics[width=10cm]{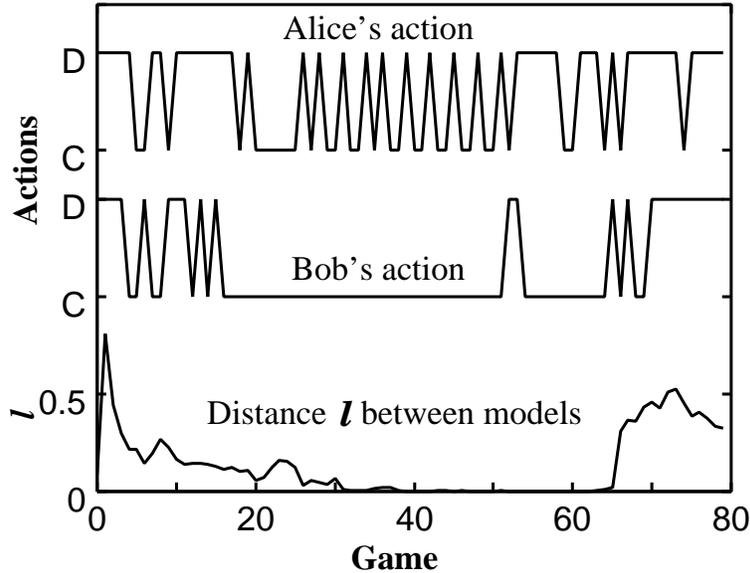}
\end{center}
\caption{``Distance'' between models with actions.
The distance is small during collaboration.
}
\label{fig:distance}
\end{figure}

\begin{center}
\begin{tt}
DDDDDCCDDCDDDDDDDDCDCCCCCCDCDCCDCCDCDCCD\\
DDDDCCDCCDDDCDCDCCCCCCCCCCCCCCCCCCCCCCCC\\
\medskip

CCDCCDCCDCCDCDDDDDDCCDDDCDCDDDDDDDCDDDDD\\
CCCCCCCCCCCCDDCCCCCCCCCCCDCDCCDDDDDDDDDD\\
\end{tt}
\end{center}

where the first and the second rows are the double-learning (Alice)
and the learning (Bob) strategies, respectively. For 50 games they can
maintain a state of collaboration. In this case, Alice had an accurate
model of Bob; therefore Alice defected as much as Bob maintained the
collaboration. However, since the model is not unique, Alice made an
inaccurate model of Bob, thus the collaboration was broken. The player using 
the double-learning strategy makes the RNN model of herself, and the
player using the learning strategy makes the RNN model of the
opponent. Therefore, the similarity between these models is a good
measure for the correctness of the model in Alice. We define the
semantical ``distance'' $l$ between models by

\begin{equation}
l = 2^{-n} \sum_{x}
(\mbox{Output}_{\mbox{\small Alice}}(x) - \mbox{Output}_{\mbox{\small Bob}}(x))^2,
\end{equation}

where the summation on $x$ is taken for all possible actions of
length $n$, and $\mbox{Output}_{\mbox{\small Alice, Bob}}$ are the outputs
from the RNN models of Alice in Alice and Bob, respectively.  Figure
\ref{fig:distance} shows the calculated ``distance'' for the same
simulation.

As expected, while Alice and Bob imagined a similar RNN model of
Alice, they could maintain the collaboration. Alice could defect so
that Bob could keep collaborating.  When they started to build the
semantically different models, Alice could not make accurate
estimations of Bob's behavior. Then when Bob started to defect, the
collaboration broke. The coincidence of the models could not be
achieved until both models entered the state of all-defection.  Many
models are able to express a particular behavior, and the selection of
the model is rather random because of the rugged landscape in the
model space. Thus, players happen to choose models of a different
behavior depending on the initial value of the RNN parameters, though
they learn the same behavior.  The non-uniqueness of models plays an
important role in this situation also.

\subsection{Discussion}
How can mutual cooperation be imported into our system?
If a player assumes
by learning that the opponent is an All-C player, he never cooperates. 
On the other hand, if he assumes that the opponent is an All-D player,
he again never cooperates. One resolution is to make player himself 
believe that the opponent is a Tit-for-Tat player. Once both
players come to believe that their opponent is TfT, they never defect
again. If the internal model never changes with the addition of the action
sequences generated by the model itself, we call it a stable model. This
self-consistent equation of the internal model dynamics determines
whether a TfT model can be stable or not.  As far as we have tested,
the attraction basin volume of TfT (i.e., the  initial set of action
sequences) hardly exists. We have also studied the meta-learning
model (i.e., to learn the internal model which the opponent learns) to
show that it is still difficult to have TfT fixed points.

There are several ways to remedy this malice situation.

 If we use finite state automata instead of recurrent neural nets, we
succeed in stabilizing several cooperating states \cite{morimoto}. The
difference arises from the truncation of learning processes in the
latter case. Namely, a player can stop learning the opponent when he
has obtained the near perfect model.  As a result, the players'
internal model dynamics have far more fixed points than does the
present case. More interestingly, several cooperative states are fixed
by mutual misunderstanding.  This truncation method can also be
applied to the present recurrent net case.

Another different approach is to relax the optimization process.
  In the present case, players make the opponent's model from the
finite history of their moves. This causes a common generalization
problem. Pollack has shown that his dynamical recognizer failed to
generalize some complex finite automaton algorithms
\cite{Pollack91}.  The generalizability of such a dynamical recognizer
should be studied further \cite{sato}.  Additionally, we encounter a
more serious problem when we study the coupled RNN.  In the case of
two Bob (or Alice) players, it is essentially impossible to learn the
other player's behavior because players cannot predict the
meta-strategic aspect of each other (e.g.  what the other player is
trying to optimize). Therefore a learning process itself loses its
meaning here.  But we see that this serious problem also arises in the
general learning process among cognitive agents \cite{bray}.  And this
we think is an example of the inevitable demon in the cognitive process.
We call it demon naming after Maxwell's demon, since it can produce
structures out of randomness in cognitive processes.

By taking such an inevitable demon seriously, we have succeeded in
achieving mutual cooperation among players of the IPD
game\cite{timt}. Namely, we let players select the internal model of
the opponent only within a certain accuracy.  We then find that orbits
of game actions, which go to mutually cooperative attractors, are
embedded in complex manners around the ``true orbit'' which goes to
the mutually defecting attractor.  As we have expected, mutually
cooperative outcomes can be established when each player comes
to assume that the other is also a Tit-for-Tat player.

There have been several attempts to make the IPD game\cite{Axelrod3}
more faithful to realistic situations. For example, a noisy IPD game
has been extensively studied. Contrary to his or her own decisions, a
player makes erroneous actions due to the influence of noise. The
purpose of such noise is to destabilize mutually-cooperating states by
creating distrust between players.  For example, two TfTs will
alternatively defect against the other.  TfT will thus be replaced by
strategies which are more tolerant of
defections\cite{Molander,Bendor,Mu}, leading further to more complex
strategies\cite{Lindgren,Ikegami}. Namely, noise brings diversity in a
society.  However, the source of noise itself has not been discussed.

A recent study on noiseless 3-person IPD games\cite{matsu} reports
that the third person will have the effect of a noise source on the
other players.  Therefore, a many-person IPD game is an another
extension of our current study.  When we extend the present analysis
to many-person games, each player should generate the image of not
only one opponent but of the relationship among players. Essential
social phenomena such as coalition, threats, altruism, etc. can thus
be discussed as emergent behavior. We hope that languages and the
abstraction of games can be observed as side effects of the evolution
of social structures. Through the present simulation, we understand
that the IPD game is a poor interactive way to achieve social
norms. We should step back and reconsider out conception of a relevant
game which incorporates social rules and structures.

\section{Summary}

The society of cognitive agents were analyzed. These agents play the
IPD games by constructing internal images based on the other's
behavior. In the case of games with fixed strategies like Tit-for-Tat
or Tit-for-Two-Tat, sound internal images were created and optimal
actions were selected. In games between cognitive agents, complex
transients were observed, and the system ultimately reached the trivial
fixed point of all defection. Complex transients emerge because of
impossibility of complete learning and the rugged landscapes in the model
space. The players change the other's strategy through their own
actions until the stable equilibrium of all defection.  These
results indicate that it is difficult to maintain mutual cooperation
in the IPD game without extra norms or communication in addition to the 
actions themselves.  It will be interesting to study whether the norms
emerge in non-zero-sum games involving the more complex actions of many
players.

\section*{Acknowledgments}

We thank Mr. Gentaro Morimoto for fruitful discussions.  This work was
partially supported by Grant-in-aid (No. 07243102) for Scientific
Research on Priority as ``System Theory of Function Emergence'' and by
Grant-in-aid (No. 09640454) from the Ministry of Education, Science,
Sports and Culture.

\end{document}